\documentclass[10pt,journal,letterpaper]{IMSTemplate}

\usepackage{times,amsmath,epsfig,subfigure,amssymb,fit,color,setspace}

\title{Parallel Matrix-Free Implementation of Frequency-Domain ~~~~~~~Finite Difference Methods for Cluster Computing}%
\author{%
Amir Geranmayeh
\vspace{12pt}\\
Continental Automotive GmbH (I ID RD EE EL ED), VDO-Str. 1, 64832
Babenhausen, Germany \\ E-mail:
amir.geranmayeh@continental-corporation.com}

\begin{document}
\maketitle


\begin{abstract}
Full-wave 3D electromagnetic simulations of complex planar devices,
multilayer interconnects, and chip packages are presented for
wide-band frequency-domain analysis using the finite difference
integration technique developed in the PETSc software package.
Initial reordering of the index assignment to the unknowns makes the
resulting system matrix diagonally dominant. The rearrangement also
facilitates the decomposition of large domain into slices for
passing the mesh information to different machines. Matrix-free
methods are then exploited to minimize the number of element-wise
multiplications and memory requirements in the construction of the
system of linear equations. Besides, the recipes provide extreme
ease of modifications in the kernel of the code. The applicability
of different Krylov subspace solvers is investigated. The accuracy
is checked through comparisons with CST MICROWAVE STUDIO$^\circledR$
transient solver results. The parallel execution of the compiled
code on specific number of processors in multi-core
distributed-memory architectures demonstrate high scalability of the
computational algorithm.

\end{abstract}


\section{Introduction}

The integrated circuits (IC) industry has previously practiced the
extraction of equivalent lumped-element models as a simulation tool
for hardware design. Advances in semiconductor interconnect
technology, such as ever increasing clock frequency and package
density, however, demand full-wave analysis of a product-level model
to preserve the signal integrity in high-speed electronic devices
\cite{Lee11}, \cite{Mittra11}. As a general multi-grid difference-based
electromagnetic field solver tool \cite{Lavranos09}, \cite{Gdansk01}, the finite integration technique
(FIT) provides a direct discretization of the Maxwell's equations in
their integral form using voltages and fluxes, respectively, along
the edges and surfaces of a pair of interlaced structured grids \cite{Schuhmann01}, \cite{Erion07}. 

To perform matrix-vector products for the matrices that do not fit in the cache, CPU requires more time to calling the matrix entries from the main memory than performing the floating-point multiplications and additions. Matrix-free methods, instead of saving the matrix in memory, recompute the operator representing matrix elements with additional floating-point operations \cite{Ren13}. On the other hand, the matrix-free methods are not able to use black-box preconditioners.

Modern object-oriented programming paradigms supply an enormous
level of abstraction in mechanisms needed within parallel computing
\cite{Parallel12}. The software package PETSc (Portable, Extensible
Toolkit for Scientific Computation) consists of an expanding set of
data structures and routines such as parallel matrix and vector
assembly which provide building blocks for facilitating the
development of scientific application codes on parallel (and serial)
computers \cite{PETSc}. PETSc uses the MPI standard for
message-passing communications and it provides a collection of
Krylov subspace (KSP) methods which can be employed as parallel
linear solvers for the frequency-domain FIT matrix systems. This
paper works out how the PETSc shell matrix concept can create a
foundation for building and solving large-scale FIT systems on
parallel processors. It is shown that the computational speed and memory demands of 3D finite-difference frequency-domain (FDFD)-based electromagnetic solvers can be scaled down simultaneously using the advanced parallel programming tools. 

The remainder of this report is organized as follows. In Section II,
the FIT discretization scheme is briefly explained. Section III
includes implementation aspects of the matrix-free-based
electromagnetic-field solver. In Section IV, the accuracy of the
developed code is checked through eigenmode calculations in a
rectangular cavity as well as comparisons with CST MICROWAVE
STUDIO$^\circledR$ (MWS) discrete-port results \cite{MWS}. The
overall performance of the parallelized algorithm is measured on
cluster of computers. Numerical simulations of complex planar
devices, multilayer interconnects, and chip packages are presented.

\section{Finite Integration Technique}\label{Sec2}

The FIT renders a consistent transformation of the integral form of
Maxwell's equations into a set of matrix equations with a unique
solution \cite{Schuhmann01}. Maxwell's equations are hereby
discretized on a pair of spatially staggered grids, namely the
primal and dual grid $(G, \widetilde G)$ with edge lengths $(L_n,
\widetilde L_n)$ and facet surface areas $(A_n, \widetilde A_n)$,
respectively, where the dual cell gridpoints $\widetilde G$ are defined on the barycenter of the primary cell mesh $G$ \cite{Clemens01}. The state variables of the FIT are the electric and
magnetic grid voltages
\begin{eqnarray}\label{eh}
\ve_n=\int_{L_n}{\bf E}({\bf r},t)\cdot {\rm d}{\bf l},\quad\vh_n=\int_{\widetilde L_n}{\bf H}({\bf r},t)\cdot {\rm d}{\bf l},\nonumber
\end{eqnarray}
the electric and magnetic grid fluxes
\begin{eqnarray}\label{db}
\fd_n=\int_{\widetilde A_n}{\bf D}({\bf r},t)\cdot {\rm d}{\bf A},\quad\fb_n=\int_{A_n}{\bf B}({\bf r},t)\cdot {\rm d}{\bf A},\nonumber
\end{eqnarray}
passing through surfaces of the pair of interlaced structured grids,
as well as the current
\begin{eqnarray}\label{Jn}
\fj_n=\int_{\widetilde A_n}{\bf J}({\bf r},t)\cdot {\rm d}{\bf A}.\nonumber
\end{eqnarray}
Storing the discrete components in vectors $\vbe$, $\vbh$, $\fbd$,
$\fbb$, and $\fbj$, the Faraday's and Amp\`{e}re's laws can be
transformed into a set of grid equations
\begin{eqnarray}\label{Maxwell}
{\bf C}\vbe=-\frac{\rm d}{{\rm d}t}\fbb,&&\quad\widetilde{\bf C}\vbh=\frac{\rm d}{{\rm d}t}\fbd+\fbj
\end{eqnarray}
in which the matrices ${\bf C}$ and $\widetilde{\bf C}$ are,
respectively, the discrete curl-operators on the primal and dual
grid. Owing to the topological nature, $\widetilde{\bf C}={\bf
C}^{\rm T}$ where ${\rm T}$ denotes the transpose of matrix. In
Cartesian grid systems with $N_p=N_x\times N_y\times N_z$ primal
grid nodes
\begin{eqnarray}\label{CP}
{\bf C}=\left(\begin{array}{ll}
~{\bf 0}\quad-{\bf P}_z\quad~{\bf P}_y\\
~{\bf P}_z\quad~{\bf 0}\quad-{\bf P}_x\\
-{\bf P}_y\quad{\bf P}_x\quad~{\bf 0}
\end{array}\right)
\end{eqnarray}
where $[P_w]_{N_p\times N_p}$ is the discrete partial
differentiation operator along $w: x, y, z$ directions and it
contains only two non-zero entries per line, $+1$ and $-1$ . The
voltages and fluxes are coupled by the diagonal material matrices
\begin{eqnarray}\label{Material}
\fbd={\bf M}_{\varepsilon}\vbe,\qquad
\vbh={\bf M}_{\mu}^{-1}\fbb.
\end{eqnarray}
The material matrices hold the locally averaged (inhomogeneous) permittivity and
permeability distribution as well as the metrics of the non-equidistant grid. The permittivity matrix elements are obtained by averaging on the dual facet $\widetilde A_n$ and the permeability matrix elements are obtained by averaging along the dual edge $\widetilde L_n$. They may consist of complex values in frequency domain to represent the dielectric or magnetic losses \cite{Rumpf14}. Enforcing the Laplace transform to the time-domain representation of Maxwell's equations in (\ref{Maxwell}), all the variables are transferred to the spectral domain. Replacing the time-derivative
$\frac{\rm d}{{\rm d}t}$ with the Laplace variable $s=j\omega$ lets us to 
analyse each frequency $f=\omega/{2\pi}$ in the time-harmonic regime individualy. Eliminating for example
$\vbh$ in (\ref{Maxwell}), the so-called wave equation w.r.t. $\vbe$
is obtained
\begin{eqnarray}\label{e2}
\begin{array}{ll}
{\bf C}^{\rm T}{\bf M}_{\mu}^{-1}{\bf C}\vbe+s^2{\bf M}_{\varepsilon}\vbe=-s\fbj
\end{array}
\end{eqnarray}
with $n_e\approx3N_p$ unknowns. The curl-curl system in (\ref{e2})
can be symmetrized by the change of variable $\vbep={\bf
M}_{\varepsilon}^{1/2}\vbe$,
\begin{eqnarray}\label{e2s}
\begin{array}{ll}
\underbrace{{\bf M}_{\varepsilon}^{-1/2}{\bf C}^{\rm T}{\bf M}_{\mu}^{-1}{\bf C}{\bf M}_{\varepsilon}^{-1/2}}_{\bf A}\vbep+s^2\vbep=-s{\bf M}_{\varepsilon}^{-1/2}\fbj
\end{array}
\end{eqnarray}
where ${\bf M}_{\varepsilon}^{-1/2}$ represents the root of inverse permittivity values and the matrix ${\bf A}$ has a block-banded structure with only 13
non-zero entries per line. Enforcing the electric boundary condition
(BC), the tangential electric field ${\bf E}_{\rm tan}$ as well as
the normal component of the magnetic flux density ${\bf B_n}$ vanish
on the perfectly electric conducting (PEC) walls.

\begin{figure}[htbp]
    \begin{center}
    \subfigure[]{\includegraphics*[width=0.42\columnwidth]{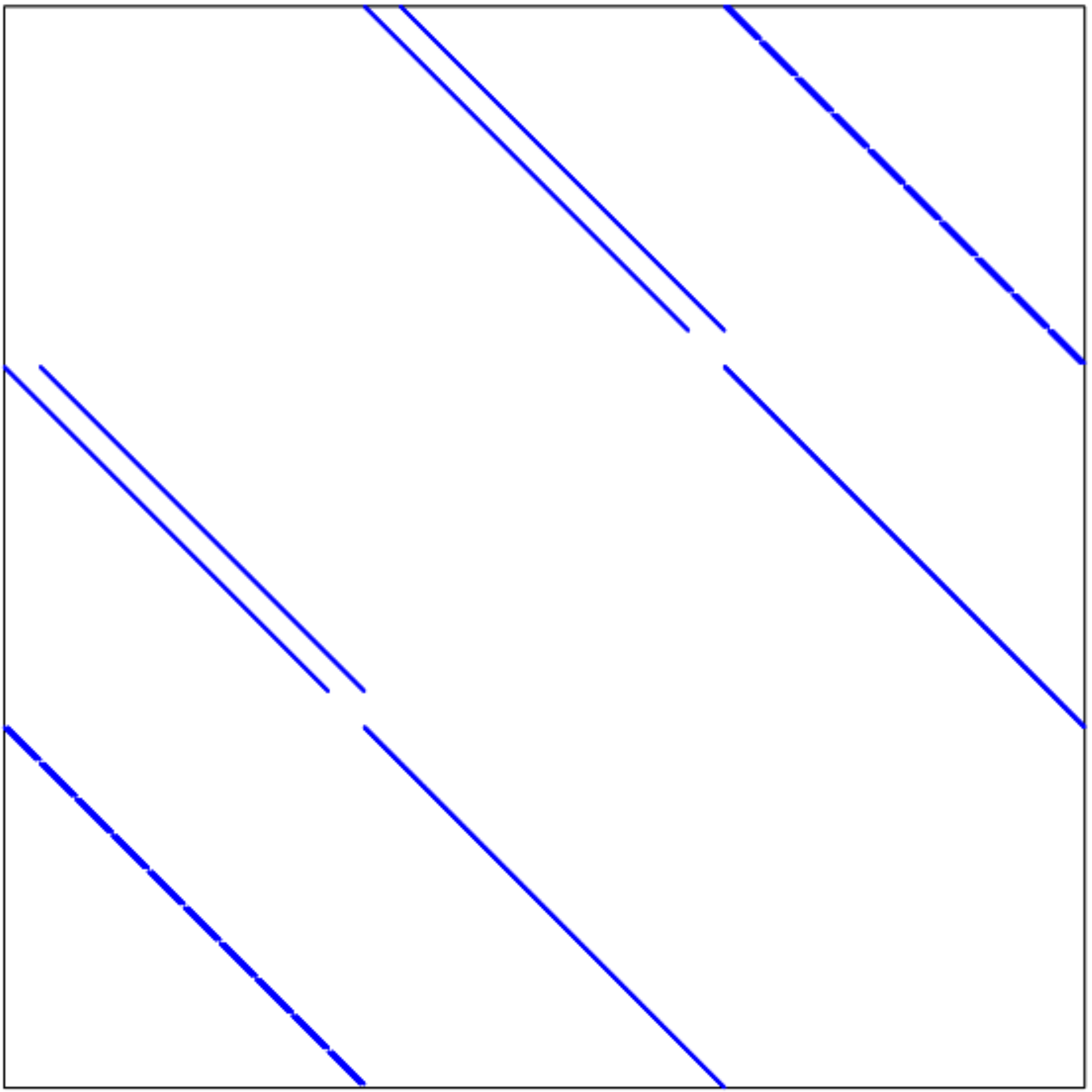}\label{SpyC}}
    \subfigure[]{\includegraphics*[width=0.42\columnwidth]{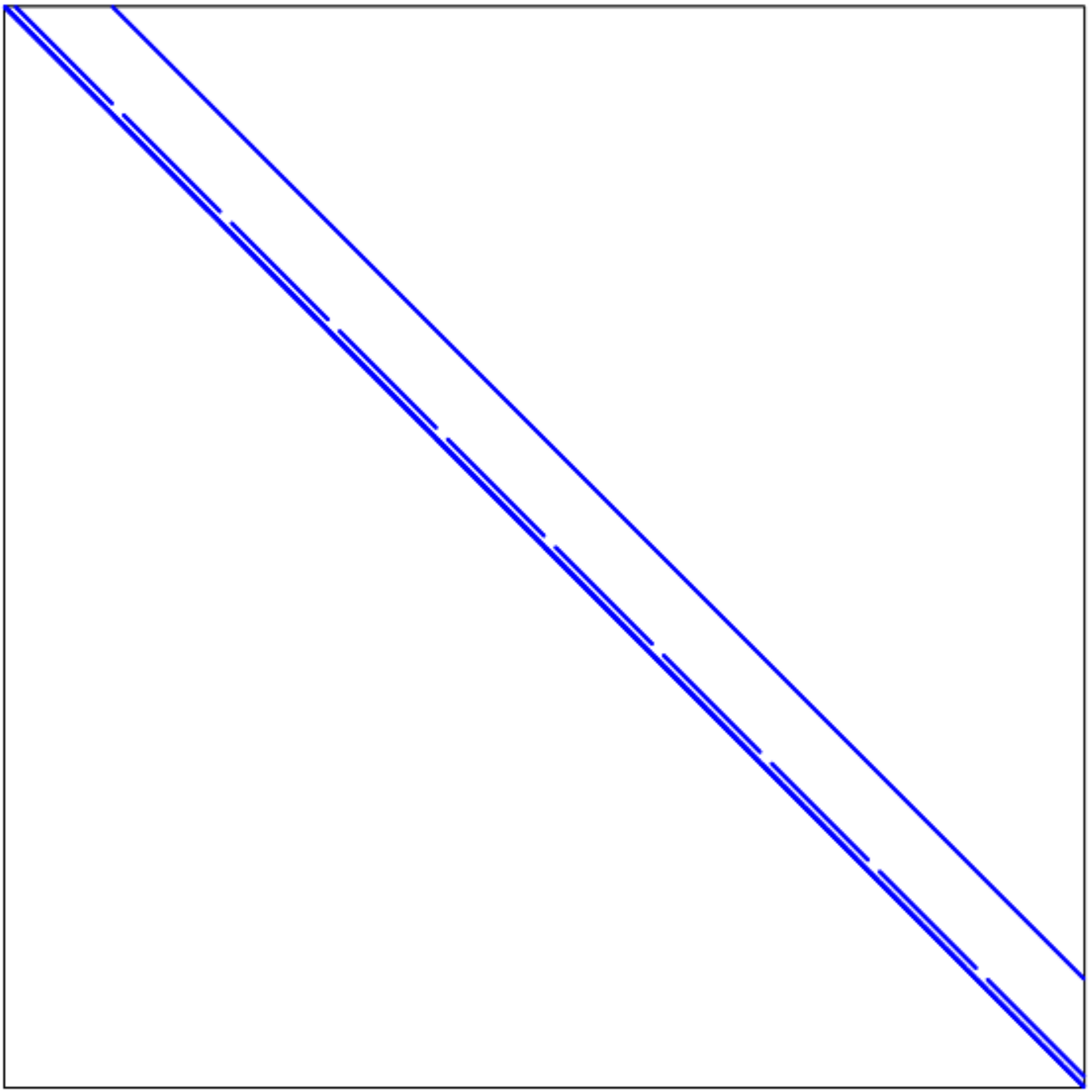}\label{SpyCxyz}}
    \end{center}
\caption{Sparsity pattern of the discrete curl matrix ${\bf C}$ for a $15\times10\times10$ grid. (a) The classical FIT (\ref{CP}). (b) The reordered numbering arrangement according to (\ref{e12Nxyz}).}\label{Spy}
\end{figure}

To accelerate the solution of the final system of equations by parallel
processing, the resulting coefficient matrix should have a
band-diagonal structure to minimize communication between multi-processors with shared or distributed memory. In order to
obtain a diagonally dominant matrix ${\bf A}$, instead of applying
the discrete derivative operator to first $x$, then $y$, and finally
$z$ components of all the grid nodes, every three in-cell $x, y, z$
unknown components are indexed one after the other cell by cell. In other words, the unknown vector components are initially arranged starting from the one with the lowest cell indices in $x$-, $y$-, $z$-directions and proceed to the higher cell indices sequentially. This corresponds to rearrange the vector components of the classical
FIT in the following manner
\begin{eqnarray}\label{e12Nxyz}
\left(\begin{array}{ll}
{\ve_1}_x\\
{\ve_2}_x\\
~\vdots\\
{\ve_{\tiny\!N}}_x\\
{\ve_1}_y\\
{\ve_2}_y\\
~\vdots\\
{\ve_{\tiny\!N}}_y\\
{\ve_1}_z\\
{\ve_2}_z\\
~\vdots\\
{\ve_{\tiny\!N}}_z
\end{array}\!\!\right)
\longrightarrow
\left(\begin{array}{ll}
{\ve_1}_x\\
{\ve_1}_y\\
{\ve_1}_z\\
{\ve_2}_x\\
{\ve_2}_y\\
{\ve_2}_z\\
~\vdots\\
~\vdots\\
~\vdots\\
{\ve_{\tiny\!N}}_x\\
{\ve_{\tiny\!N}}_y\\
{\ve_{\tiny\!N}}_z
\end{array}\!\!\right).
\end{eqnarray}
This standard cache-based architecture is used for the layout of unknowns in PETSc. Of course, to avoid any matrix permutation, the above index
assignment are followed from the early stage of the discretization,
Fig.~\ref{Spy}. 

The complete-solution procedure briefly involves the following
successive numerical stages: preparation of the (possibly complex)
permittivity and permeability matrices, construction of the discrete
curl matrix, incorporation of the boundary conditions to the
material matrices, setting up the excitation vector, assemblage of
the system matrix equation for the frequency of interest, successive
solution of the obtained sparse system of equation, and at last
post-processing of the results for monitoring the field quantities
on the desired probes.

\section{Matrix-Free Method}\label{Sec3}

The memory limitations for the storage of the system matrix and the
associated subspaces prohibit the usage of adequately fine grids for
an accurate modeling of electrically-small geometrical complexities.
As a remedy, the matrix-free methods do not require explicit storage
of the product of matrices in (\ref{e2s}). Replacing the Laplace operator $s$ in (\ref{e2s}) with the angular frequency $j\omega$, the matrix-free algorithm
requires only the application of the linear matrix operator to a
vector,
\begin{eqnarray}\label{e2t}
\begin{array}{ll}
\underbrace{\underbrace{{\bf M}_{\varepsilon}^{-1/2}\underbrace{{\bf C}^{\rm T}\underbrace{{\bf M}_{\mu}^{-1}\underbrace{{\bf C}\underbrace{{\bf M}_{\varepsilon}^{-1/2}\vbep}_{{\bf x}^{(1)}}}_{{\bf y^{(1)}}}}_{{\bf x^{(2)}}}}_{{\bf y^{(2)}}}}_{{\bf x^{(3)}}}-\omega^2{\bf I}\vbep}_{{\bf y^{(3)}}}=-j\omega{\bf M}_{\varepsilon}^{-1/2}\fbj
\end{array}
\end{eqnarray}
where ${\bf I}$ is the identity matrix and $.^{(q)}$ represents the $q^{\rm th}$ reuse of the memory buffers ${\bf x}$ and ${\bf y}$. Thus, properly user-defined
shell matrix operations can be applied instead of explicit
assemblage of ${\bf A}$, using for example the PETSc matrix
operations {\small\verb"MatMult()"} and
{\small\verb"MatMultTranspose()"} between three
{\small\verb"VecPointwiseMult()"} vector operations plus
{\small\verb"VecAXPY()"} for summing up the scalar $-\omega^2$ term.
Therefore, the matrix-free implementation of the square
matrix-vector products in FIT requires only 2 temporary vectors,
namely ${\bf x}$ and ${\bf y}$ as (\ref{e2t}) implies, and 4
temporary vectors when ${\bf C}$ is non-square due to the possible
elimination of null rows. 

For preallocation of the sparse parallel
matrix ${\bf C}$ via {\small\verb"MatCreateMPIAIJ()"}, the number of
nonzeros per row is set to 4. To store ${\bf C}$ with 4 non-zero per line when using a CSR-format, $(4+8)\,4\,n_e$ bytes of memory is required for the storage of the integer column indices and the nonzero real values. Indeed, the CSR-format also requires a third array with length $4n_e$ (in fact $n_e+1$ entries with typically 4 bytes each) pointing to the beginning of each row. This third array does affect the memory consumption here as the average number of nonzeros per row is relatively small. Additionally, two floating-point arrays, each allocating $8n_e$ bytes, are needed to store the diagonal elements of the material matrices ${\bf M}_{\varepsilon}^{-1/2}$ and ${\bf M}_{\mu}^{-1}$ in double-precision.
Hence, the matrix-free matrices in
(\ref{e2t}) demands $8\,n_e+(4+4+8)\,4\,n_e+8\,n_e=80\,n_e$ bytes of
memory to store ${\bf C}$ and diagonals of the material matrices.

The explicit sparse storage of the FIT-system matrix
${\bf A}$ with 13 double-precision entries per line in (\ref{e2s})
requires $(4+8)\,13\,n_e=156\,n_e$ bytes plus the additional
$8\,n_e$ bytes for saving ${\bf M}_{\varepsilon}^{-1/2}$ needed for
the final voltage scaling. Note that the classical FIT (\ref{e2}) or
(\ref{e2s}) entail the multiplication of 13 non-zero elements in
$n_e$ rows of the system matrix for every iteration, whereas in
(\ref{e2t}) only $(1+4+1+4+1)$ entries have to be multiplied per
row. Note that the summation of $-\omega^2$ term should be added to the operation counts of the matrix free methods. 

On the other hand, considering ${\bf A}={\cal A}^{\rm T}\!{\cal A}$
where ${\cal A}={\bf M}_{\mu}^{-1/2}{\bf C}{\bf
M}_{\varepsilon}^{-1/2}$ and ${\bf M}_{\mu}^{-1/2}$ represents the root of inverse permeability values, one can construct ${\cal A}$ and
alternatively utilize the above-mentioned two shell matrix
operations
\begin{eqnarray}\label{e2tt}
\begin{array}{ll}
\underbrace{\underbrace{({\bf M}_{\mu}^{-1/2}{\bf C}{\bf M}_{\varepsilon}^{-1/2})^{\rm T}\underbrace{{({\bf M}_{\mu}^{-1/2}{\bf C}{\bf M}_{\varepsilon}^{-1/2})}\vbep}_{\bf t}}_{\bf x}-\omega^2{\bf I}\vbep}_{\bf y}.
\end{array}
\end{eqnarray}
This yields a more efficient algorithm than (\ref{e2t}) due to the
less vector operations and overwriting ${\cal A}$ on ${\bf C}$ by
applying {\small\verb"MatDiagonalScale()"} and deallocating the
memory reserved for ${\bf M}_{\mu}^{-1/2}$ shrinks the memory usage
to $(4+4+8)\,4\,n_e+8\,n_e=72\,n_e$ bytes. This calls for $(4+4)\,n_e$
multiplications per iteration and $2\,n_e$ multiplications for the
construction of $\cal A$ in advance.

The KSP solvers in PETSc support matrix-free methods. 
Matrix-free methods efficiently carry out matrix-vector products but
the associated iterative solvers do not work well without a
preconditioner \cite{Shin13}. Hence, a user-defined preconditioner should be
devised \cite{Tijhuis08}. An additional matrix-vector product, however, is not
necessarily needed to extract for example the diagonal elements of
the shell matrix, as a diagonal preconditioner can be constructed by
taking the sum square elements of the columns of the sparse matrix
${\cal A}$, i.e.,
\begin{eqnarray}\label{Precond}
[{\cal P}]_{ii}=\!\sum_k[{\bf M}_{\mu}^{-1}]_{kk}[{\bf C}]_{ki}^2[{\bf M}_{\varepsilon}^{-1}]_{ii}\!=\!\!\sum_{k, j=i}\![{\cal A}]_{ik}^{\rm T}[{\cal A}]_{kj}\!=\!\sum_k[{\cal A}]_{ki}^2.\nonumber
\end{eqnarray}
Table \ref{Tab:Ne} reflects the computational costs and storage
demands for the introduced formulations.
\begin{table}[!t]
\centering
    \caption{Comparison of the required element-wise multiplications per iteration and memory consumptions for different types of FIT implementations with $n_e$ degrees of freedom.}
    \label{Tab:Ne}
    \begin{small}
    \begin{tabular}{|l|l|l|l|l|}
    \hline
    { FIT } & \multicolumn{2} {c|} { Classical } & \multicolumn{2} {c|} { Matrix-Free } \\
    \hline
    { Algorithms } & ~(\ref{e2}) & ~(\ref{e2s}) & ~(\ref{e2t}) & ~(\ref{e2tt}) \\
    \cline{1-5}
    { Symmetric } & ~$\times$ & ~$\checkmark$ & ~$\checkmark$ & ~$\checkmark$ \\
    \hline
    { Memory Usage } & $164\,n_e$ & $164\,n_e$ & $80\,n_e$ & $72\,n_e$ \\
    \hline
    { Multiplications } & $13\,n_e$ & $13\,n_e$ & $12\,n_e$ & $9\,n_e$ \\
    \hline
    \end{tabular}
    \end{small}
\end{table}

When an extremely large mesh has to be generated, in
the pre-processing stage the simulation domain can simply be decomposed
into slices along the $z$-axis to accumulate the information associated with
slices of material matrices on different machines. The PETSc provides distributed arrays which can be decomposed along all three axes. The slicing along the $z$-axis is an intuitive option which facilitates the implementation and might not be the optimal domain decomposition. Other options for mesh partitioning are using the METIS or Scotch software packages. The optimal domain decomposition for the Jacoby method is the one which minimizes the surface-to-volume ratio of each submesh owned by an MPI rank \cite{Hager11}.

\section{Numerical Results and Discussion}\label{Sec4}

The correctness of the code was first checked through eigenmode
calculations of rectangular and circular cavities for which
analytical solution exists. In the following examples, time-harmonic
currents ${\bf i}=\fbj$ are imposed at the input ports as the
excitation signal and the voltages ${\bf u}={\bf
M}_{\varepsilon}^{-1/2}\vbep$ along the segments of desired probes
are summed up as the output values. The relative permeability
$\mu_r$=1 and zero-conductivity $\sigma=0\;\rm S/m$ are assumed for
the material properties, unless otherwise mentioned. A relative
decrease to less than $10^{-12}$ in the residual norm is set as the
stop criterion for iterative solvers and the internal impedance of
the source is assumed $Z_0=50\:\Omega$ for s-parameter calculations.
The obtained impedance (Z)-parameters of multi-port network devices
are compared with CST MICROWAVE STUDIO$^\circledR$ (MWS) transient
solver results with discrete face-port excitation. The CST MWS benchmark results are obtained on the same mesh grids used for the presented frequency-domain methods. The presented schemes also cover the usage of higher-order FIT for faster convergence, since only the material matrices would then be affected.

\begin{figure}[htbp]
    \centerline{\psfig{figure=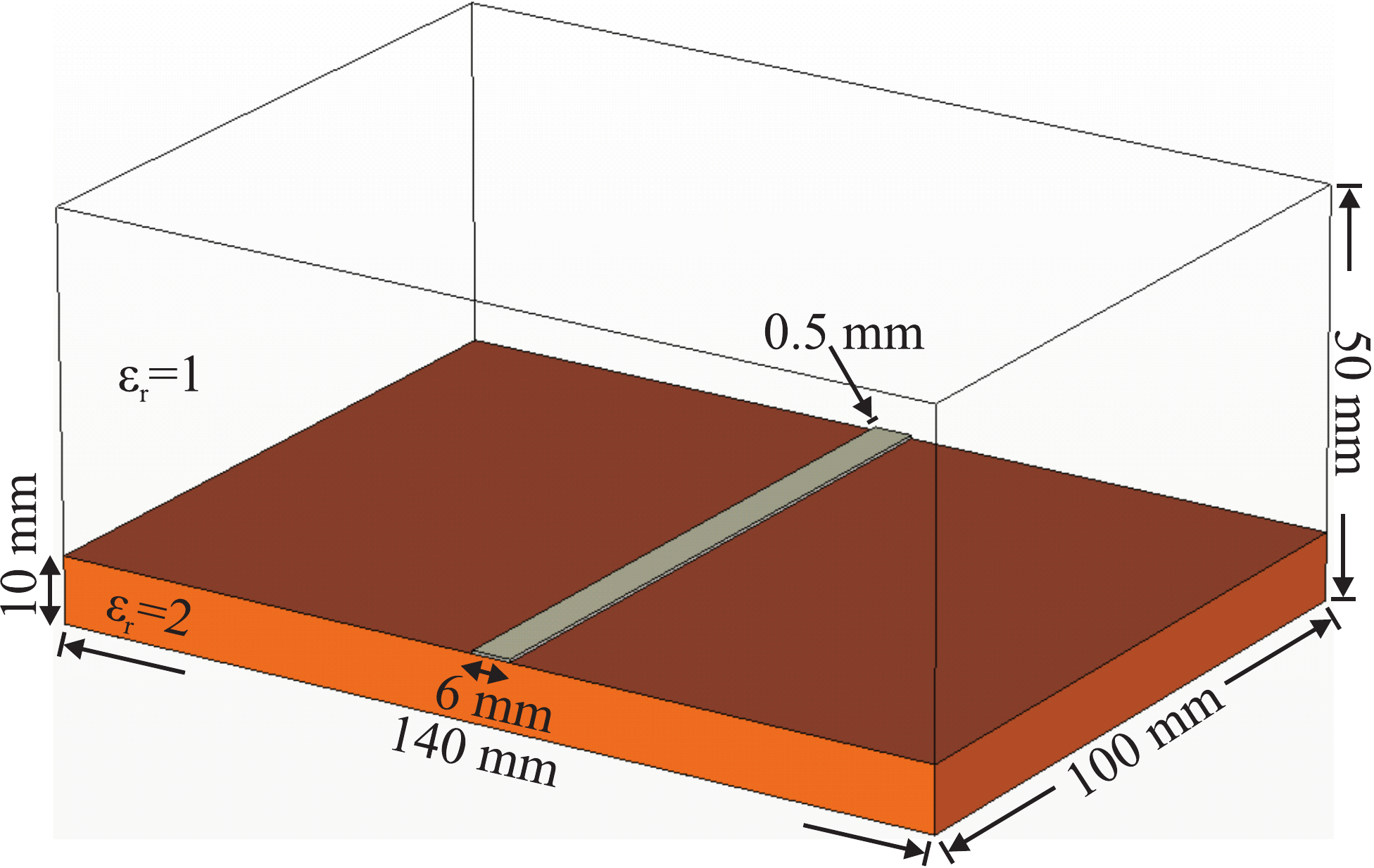,scale=0.40}}
    \caption{A microstrip line enclosed in a metallic box. For the clarity of representation, the surrounding box was represented as a transparent one.}
    \label{MicrostripLine}
\end{figure}

\begin{figure}[htbp]
    \begin{center}
    \subfigure[]{\includegraphics*[width=0.49\columnwidth]{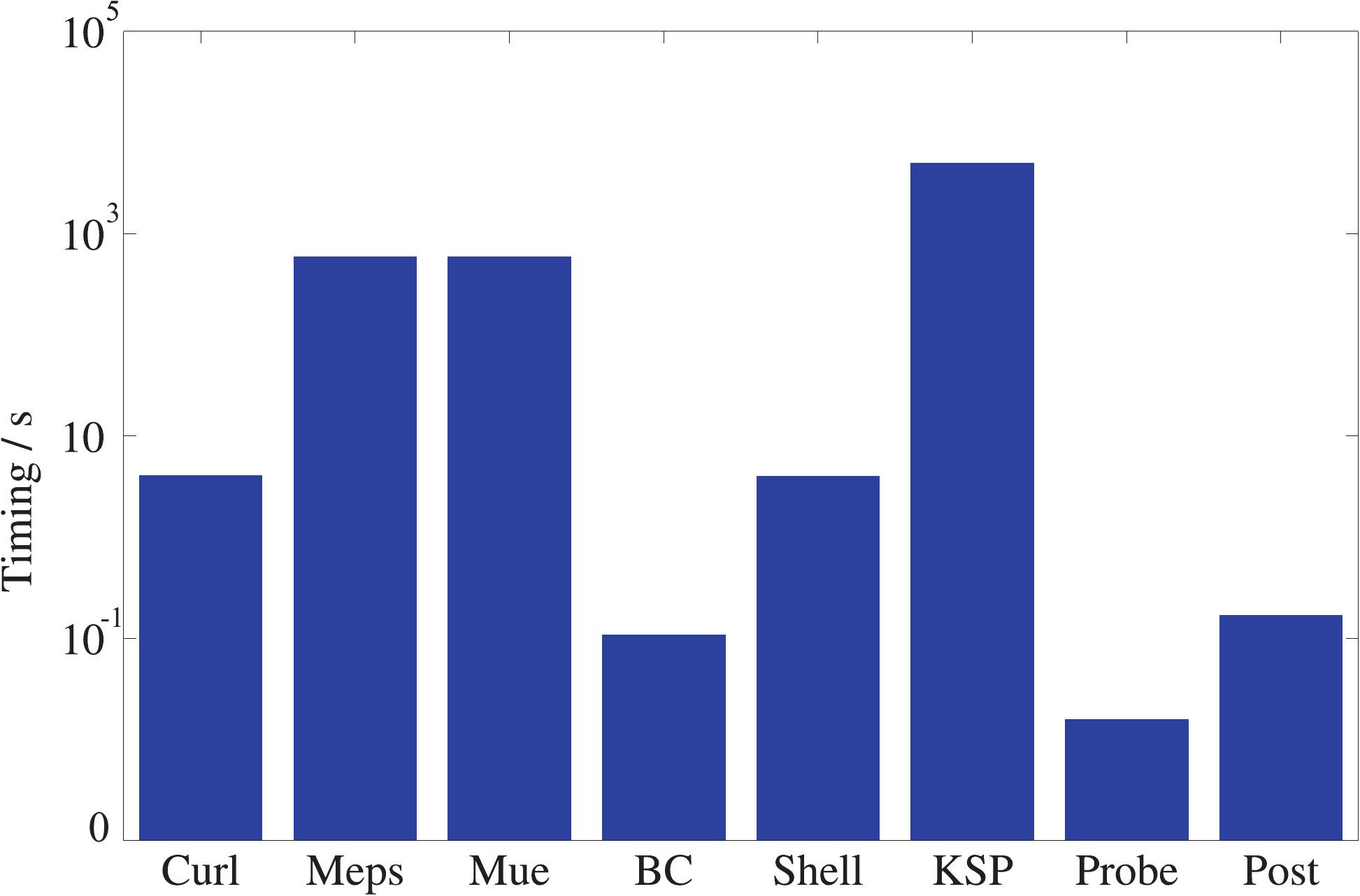}\label{Timing}}
    \subfigure[]{\includegraphics*[width=0.49\columnwidth]{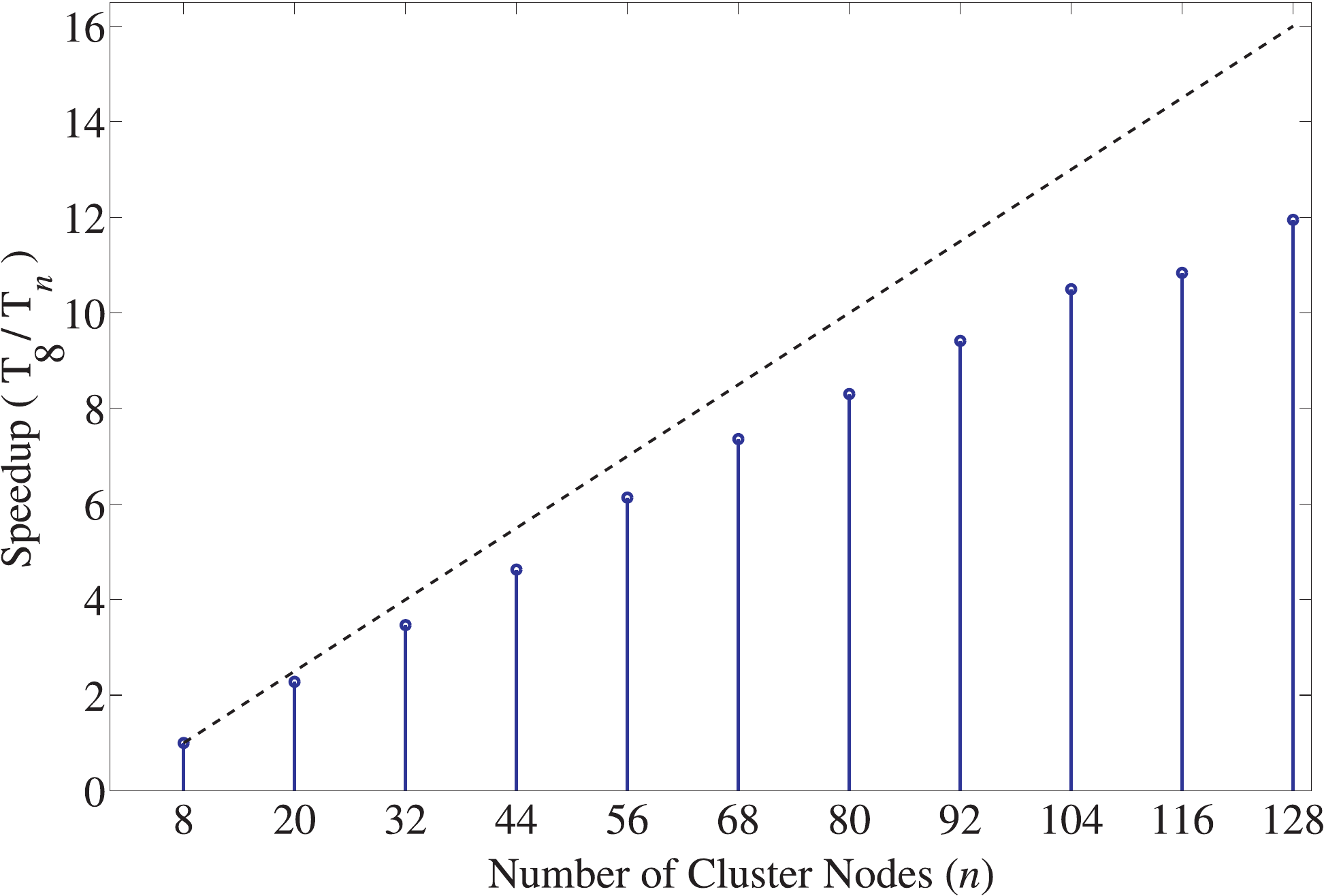}\label{Speedup}}
    \end{center}
\caption{(a) The timing of different stages of constructing (\ref{e2t}) for the problem sketched in Fig.~\ref{MicrostripLine} discretized by $849\times608\times327$ mesh cells and solving it for $n_e$=\,506\,384\,352 unknowns on 96 nodes using the conjugate gradient method. (b) The speedup diagram of solving the matrix equation resulting from (\ref{e2t}) for the microstrip line problem depicted in Fig.~\ref{MicrostripLine} discretized by $497\times354\times193$ mesh cells on clusters of the workstation using the KSP conjugate gradient method. The 8 nodes timing is used as reference.}\label{}
\end{figure}

As the first example, a microstrip line enclosed in a PEC box as depicted in Fig.~\ref{MicrostripLine} is analyzed. Fig.~{\ref{Timing}} illustrates the execution times of different blocks of the program in modeling the microstrip line at a single frequency $f=0.5$ GHz on 96 processors in a distributed memory architecture using the conjugate gradient method. The $x$-axis tick labels represent different stages of the solution process, i.e. respectively, preparation of the discrete curl, permittivity, and permeability matrices, inclusion of boundary conditions in the material matrices, stimulation of excitation ports and assemblage of the frequency-dependant shell system matrix, iterative solution of the obtained system of equations using Krylov subspace methods, determination of observation probes to store the desired part of the solution, and finally the post-processing impedance calculations. As displayed in Fig.~{\ref{Timing}}, the construction of the shell matrix does not take much time in comparison with the KSP solver iterations till convergence. 

To study the solver speedup under uniform memory access time on multi-core shared memory processsors, a quad-core 64-bit Intel Xeon X5647 CPU with 12M cache and 24 GB of RAM is considered. The
speedup factor reaches to 3 for solving the microstrip line problem
with hundred thousand unknowns on 4 cores, that is 75\% efficiency,
and it falls down to the same efficiency factor 75\% on 2 cores due to the memory bandwidth limitations for
larger problems with million unknowns.

%

\begin{figure}[htbp]
    \centerline{\psfig{figure=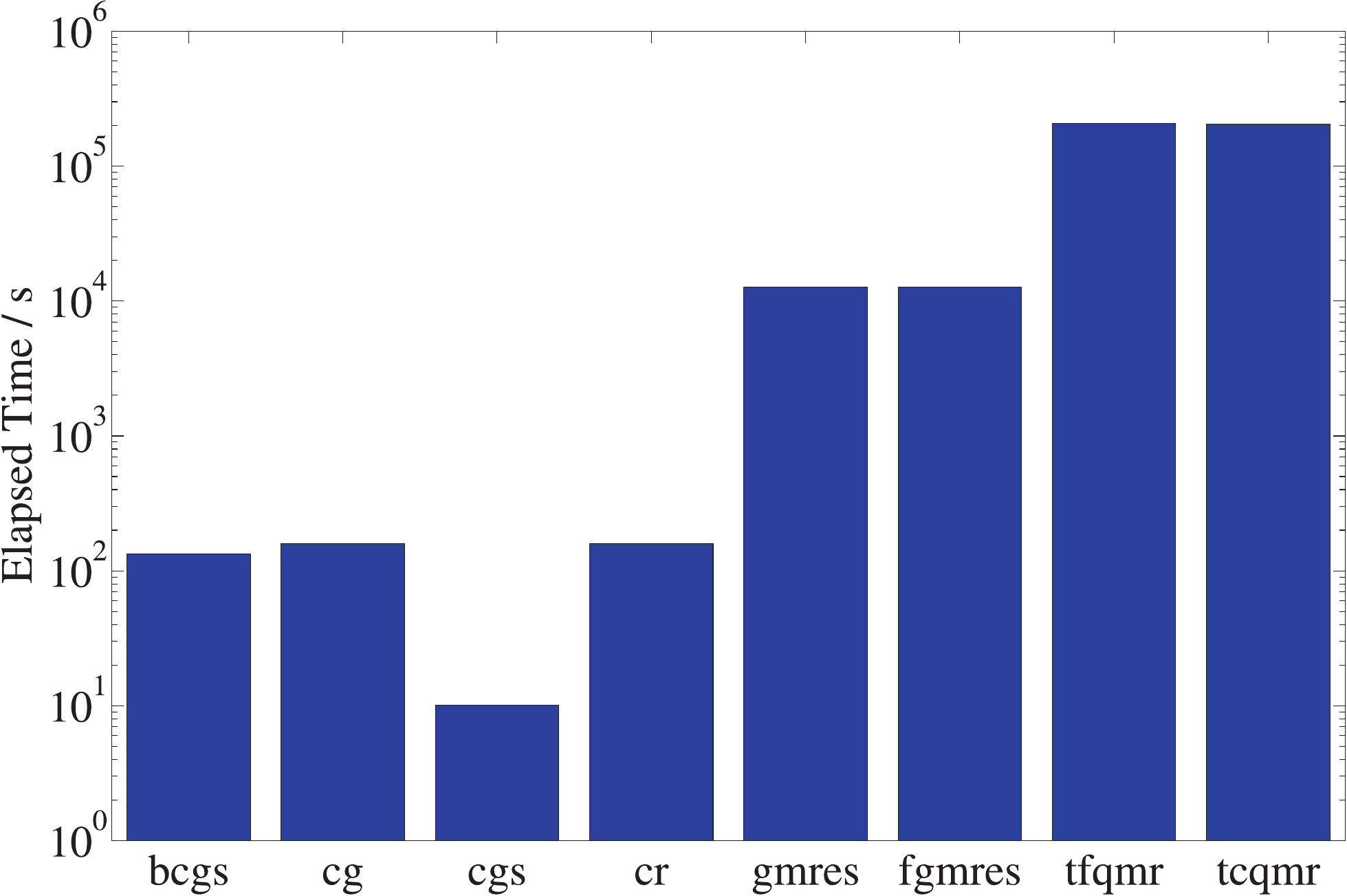,scale=0.46}}
    \caption{The solution time for different PETSc iterative methods: the biconjugate gradient stabilized (bcgs), conjugate gradient (cg), conjugate gradient squared (cgs), conjugate residual (cr), generalized minimal residual (gmres), flexible gmres (fgmres), two variants of transpose-free quasi-minimal residual (tfqmr and tcqmr) methods.}
    \label{Iterative}
\end{figure}

Fig.~{\ref{Speedup}} demonstrates the speedup factor in solving the obtained linear system of equations using the KSP conjugate gradient method. Fig.~{\ref{Iterative}} compares the elapsed CPU time for solving (\ref{e2t}) on 64 nodes for $n_e$=10\,212\,276 unknowns associated with the problem shown in Fig.~\ref{MicrostripLine} by different iterative methods provided in PETSc 3.4 \cite{PETSc}. 
Although the conjugate gradient squared (cgs) method apparently converges faster than the others, its matrix-free implementation does not deliver accurate results close to the resonant frequencies of the large-size problems, and hence, it is excluded from the rest of paper. 
The generalized minimal residual method (gmres) family have shown to be relatively slow. It seems that the explicitly stored Krylov space generated by Gram-Schmidt orthogonalisation constitutes the overhead for gmres. Note that the flexible gmres (fgmres) is equivalent to gmres as long as the preconditionner is not modified during the iterations. It is also worth noting that the conjugate gradient (cg) method is only applicable if the system matrix ${\bf A}$ is symmetric positive definite (SPD), i.e. if ${\cal A}$ defined before (\ref{e2tt}) is invertible. 

\begin{figure}[htbp]
    \begin{center}
    \subfigure[]{\includegraphics*[scale=0.41]{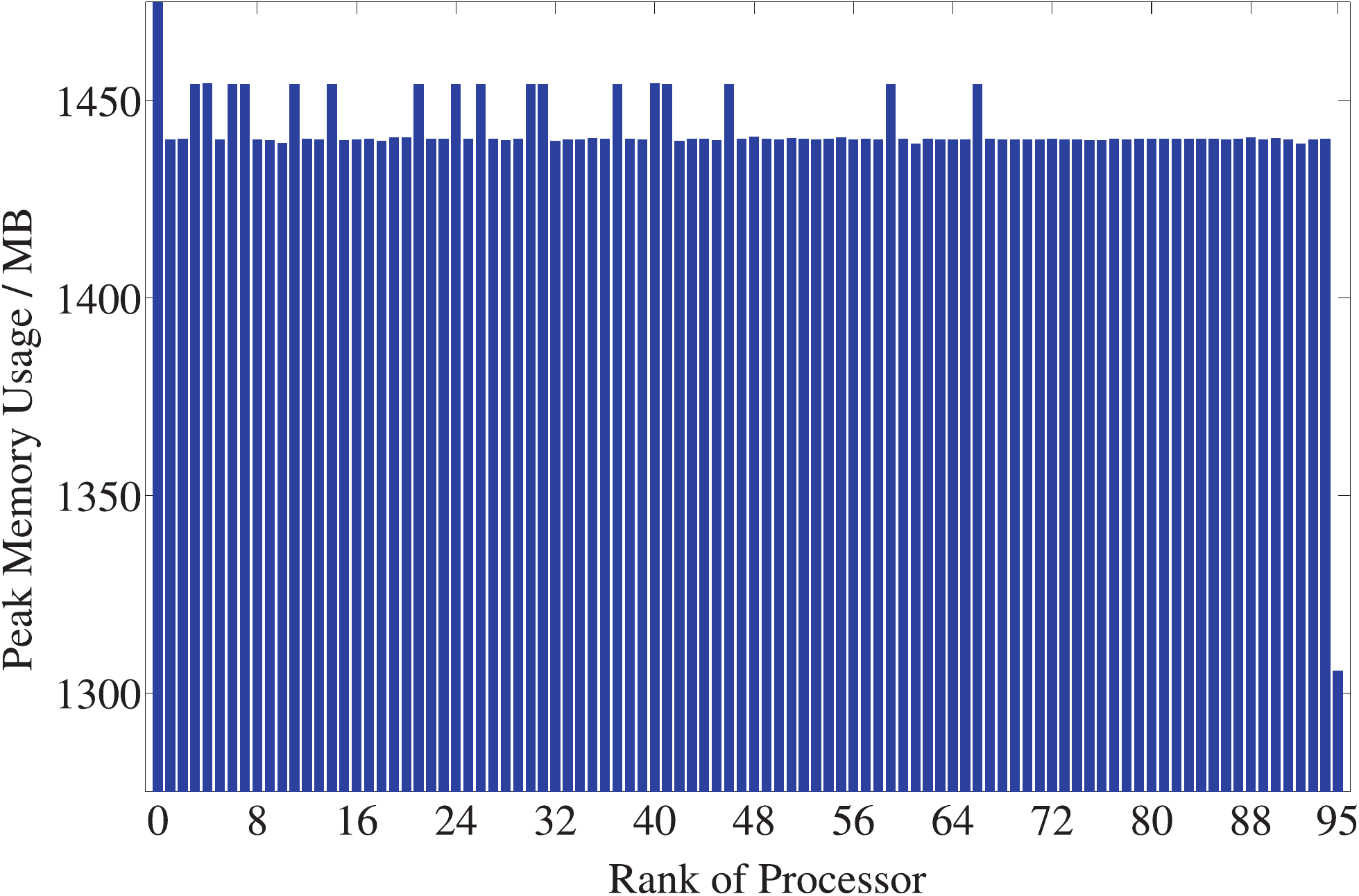}\label{LoadBalancea}}
    \subfigure[]{\includegraphics*[scale=0.41]{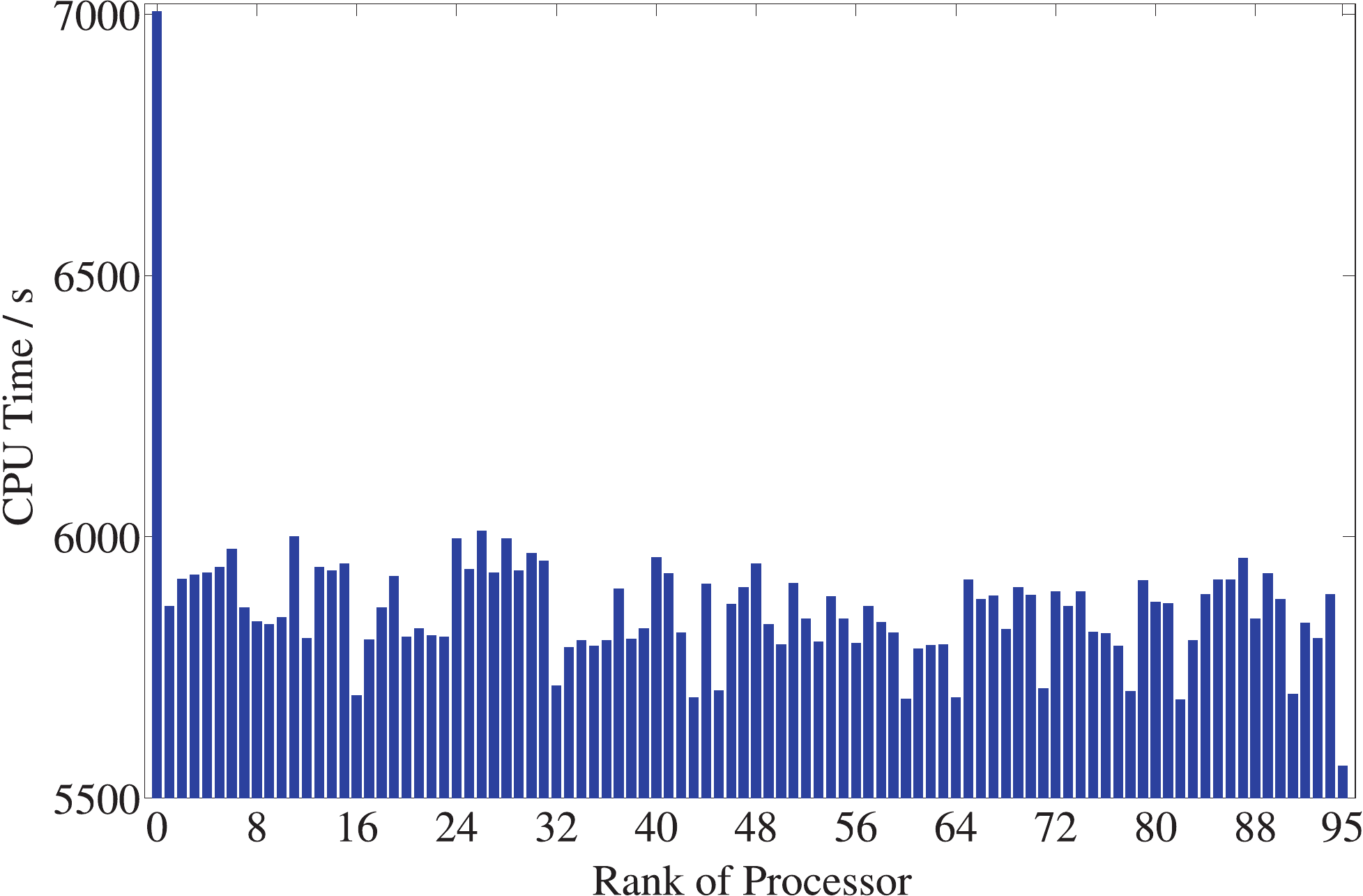}\label{LoadBalanceb}}
    \end{center}
\caption{The load-balance in simulation of the microstrip line at $f=0.5$ GHz discretized by $849\times608\times327$ mesh cells resulting in $n_e$=\,506\,384\,352 unknowns distributed on 96 nodes solved by the biconjugate gradient stabilized method.}\label{LoadBalance}
\end{figure}

\begin{figure}[htbp]
    \begin{center}
    \subfigure[]{\includegraphics*[width=0.49\columnwidth]{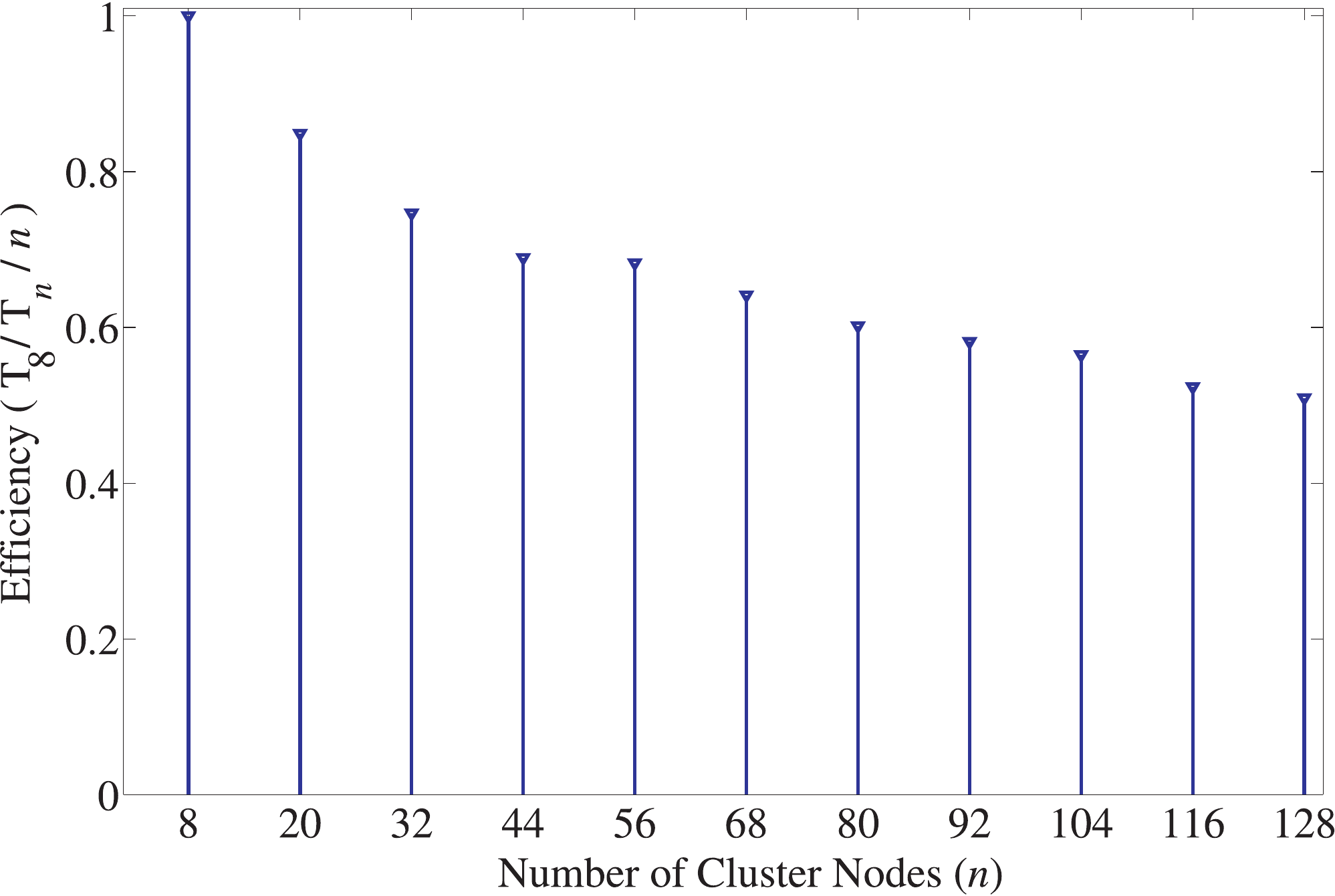}\label{Efficiency}}
    \subfigure[]{\includegraphics*[width=0.49\columnwidth]{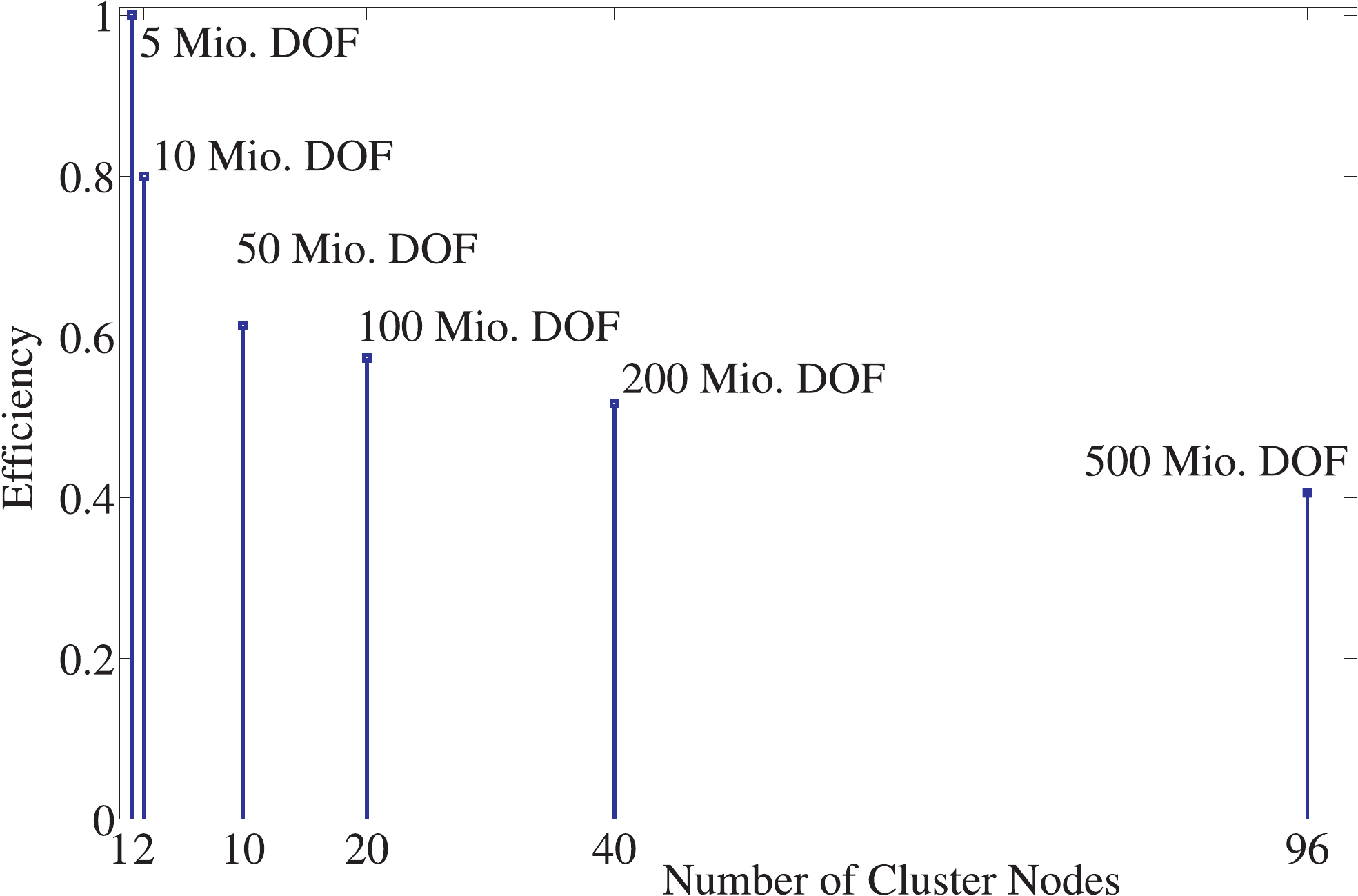}\label{EfficiencyScale}}
    \end{center}
\caption{(a) The strong scaling experiment with the simulation of the microstrip line at $f=0.5$ GHz for the fixed problem size of $n_e$=101\,868\,102 on a distributed-memory machine using the biconjugate gradient stabilized method. (b) The weak scaling experiment on the simulation of the microstrip line with the help of biconjugate gradient stabilized method for a constant ratio of workload to the used nodes in the cluster.}\label{}
\end{figure}

Fig.~{\ref{LoadBalance}} demonstrates the balance of workload among
the used 96 processors for the solution of $n_e$=506\,384\,352 FIT
unknowns using the biconjugate gradient stabilized method. Less than 0.1 second imbalance is observed in the total
run time of different processors due to I/O and less than 1.6 MB in the peak
memory consumption among the ranks. Note that the execution time and peak memory usage for the construction of material matrices on the first rank has also been included in Fig.~{\ref{LoadBalance}}. Fig.~{\ref{Efficiency}} displays the strong scaling (i.e. fixed global problem size) experiment, considering the complete solution time when the problem size
is kept fixed $n_e$=101\,868\,102 and solved by the biconjugate gradient stabilized method.
Fig.~{\ref{EfficiencyScale}} exhibits the weak scaling (i.e. fixed problem size per node) experiment, cosidering the complete solution procedure with the biconjugate gradient stabilized method when the number of unknowns are proportionally
scaled with the number of cluster nodes. The iteration counts for different system sizes grow almost linearly. 

\begin{figure}[htbp]
    \begin{center}
    \subfigure[]{\includegraphics*[width=0.49\columnwidth]{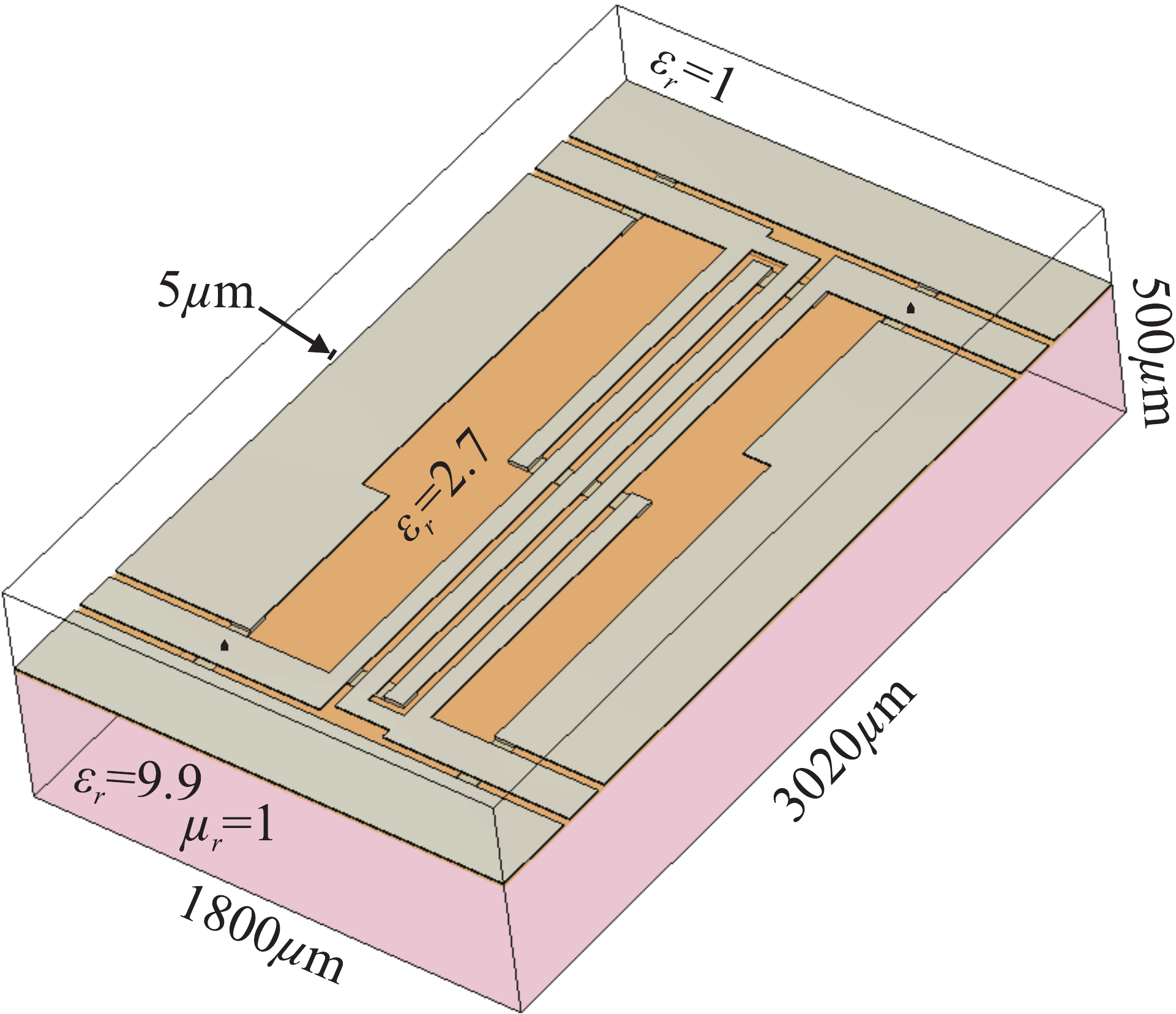}\label{CoplanarCoupler}}
    \subfigure[]{\includegraphics*[width=0.49\columnwidth]{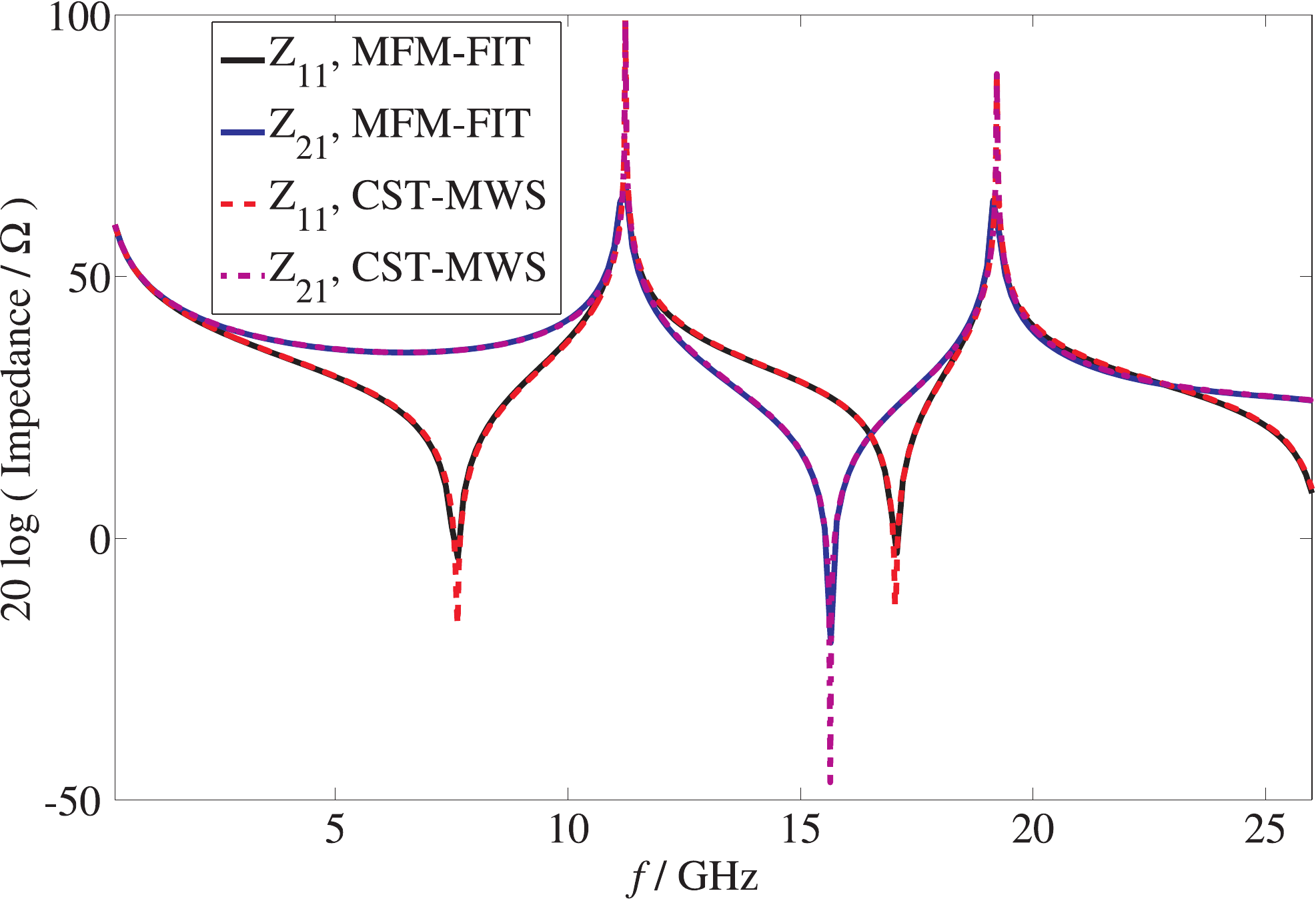}\label{Z21CoplanarCoupler}}
    \end{center}
\caption{(a) A four-port coplanar coupler without ground plane. The exact geometrical details can be found in CST-MWS examples \cite{MWS}. (b) The input impedance $Z_{11}$ and the forward gain $Z_{21}$ for the coplanar coupler shown in Fig.~\ref{CoplanarCoupler}.}\label{}
\end{figure}


The correctness of the parallelized matrix-free implementation for the four short-listed robust iterative solvers from Fig.~\ref{Iterative}, namely the bcgs, cr, cg, and gmres methods, are now investigated on four complex case studies. 
As another example of planar devices, a four-port coupler consisting of coplanar lines without ground plane planted on a dielectric substrate with the relative permittivity $\epsilon_r$=2.7 is considered. The structure is placed up on a $500~\mu$m thick aluminium layer as shown in Fig.~\ref{CoplanarCoupler}. The thickness of the thin substrate is $13~\mu$m. Ports are placed vertically between the striplines and the underlay ground to let the two (even and odd) quasi-TEM modes be able to propagate along the lines. Perfect magnetic conductors (PMC) are considered on all sides of the structure to avoid the excitation of auxiliary box modes. The structured mesh $N_x$=46, $N_y$=73, $N_z$=16 results in $n_e$=161\,184 degrees of freedom for which (\ref{e2t}) is solved at $N_f$=200 equidistant frequencies within the range $[f_{\rm min},f_{\rm max}]=[0.26,26]$ GHz using the generalized minimal residual method. As shown in Fig.~\ref{Z21CoplanarCoupler}, the matrix-free method (MFM)-FIT results agree well with the CST-MWS time-domain solver results on the same hexahedral mesh. Note that, the present implementation is equipped with the perfect boundary approximation (PBA)$^\circledR$ technique for accurate modeling of curved objects. 

\begin{figure}[htbp]
    \begin{center}
    \subfigure[]{\includegraphics*[width=0.49\columnwidth]{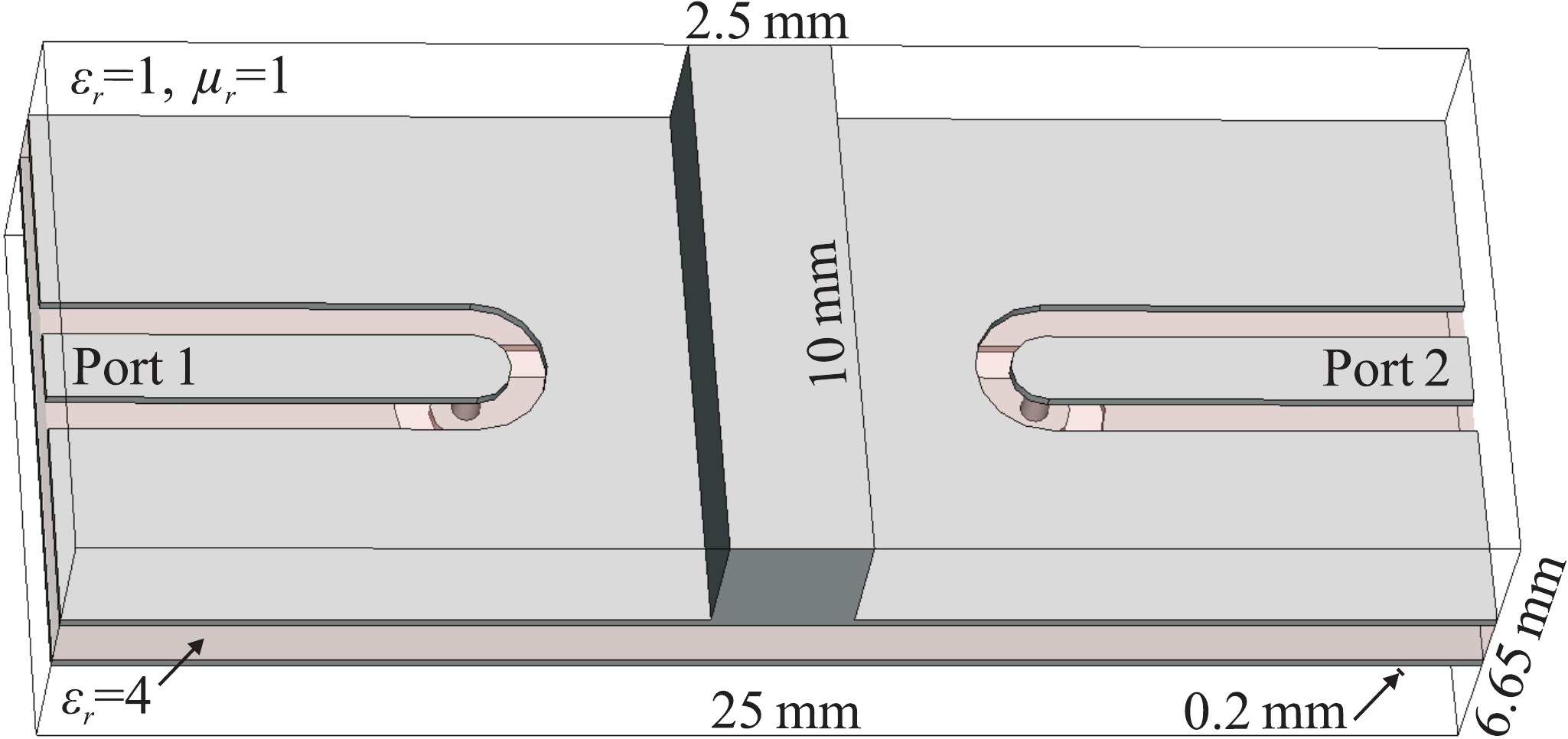}\label{Trans}}
    \subfigure[]{\includegraphics*[width=0.49\columnwidth]{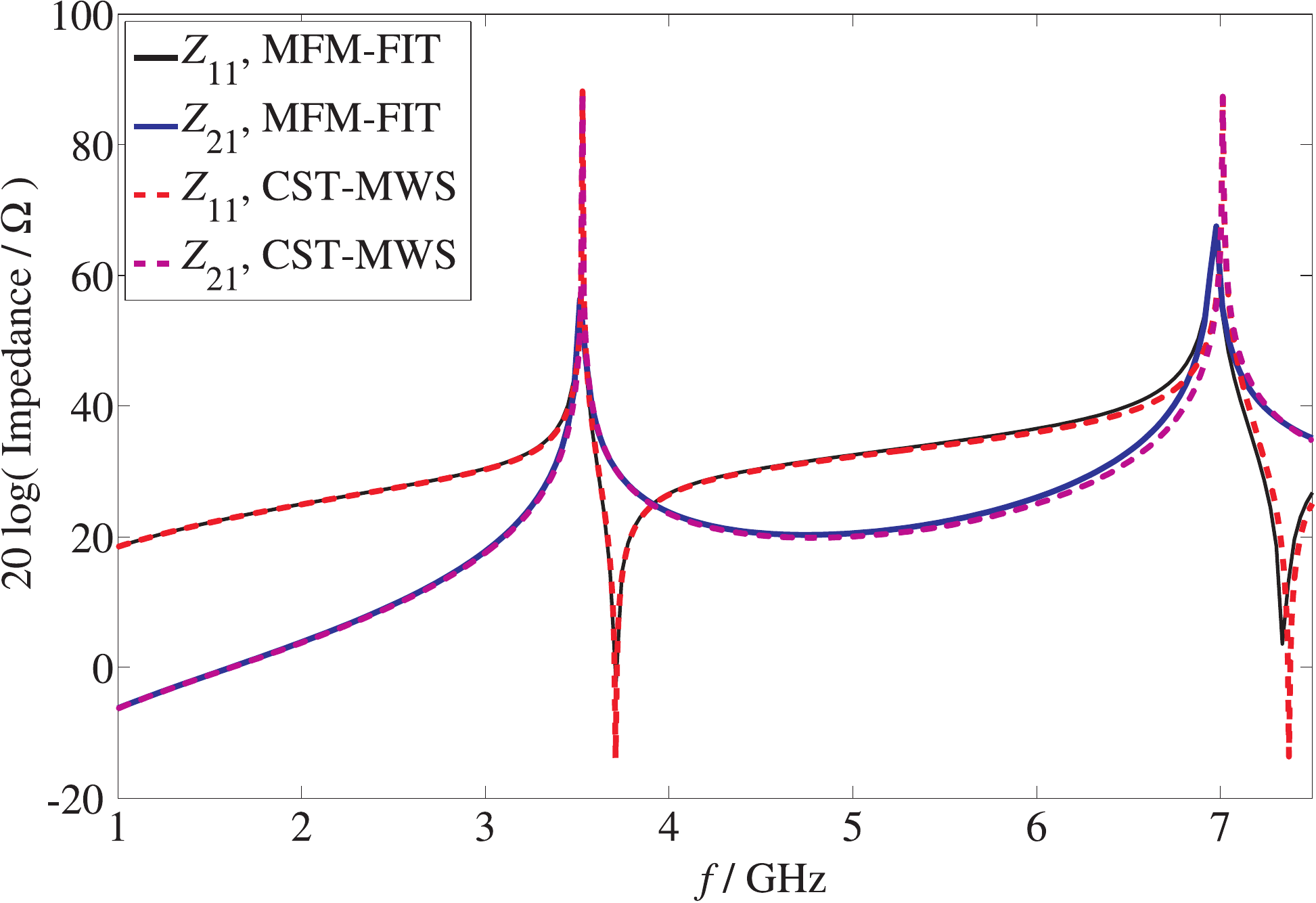}\label{Z21Trans}}
    \end{center}
\caption{(a) A transmission line using strip lines in three dimensions placed on a dielectric substrate. The strip lines on the top surface of the substrate are connected to a short strip line on the bottom surface by small cylinders enabling the bypass of a possible barrier. (b) The input impedance $Z_{11}$ and the forward gain $Z_{21}$ for the microstrip transmission line in Fig.~\ref{Trans}.}\label{}
\end{figure}

The next structure to be analyzed contains two vias for transmitting
the signal from one side of the printed circuit board (PCB) to the
other side while bridging over a metal barrier, as depicted in
Fig.~\ref{Trans}. The thickness of the substrate is $1.25$ mm. The
PEC boundary condition is considered for all sides of the structure,
except the top and bottom sides which are terminated by PMC. The
hexahedral mesh $N_x=31$, $N_y=77$, $N_z=22$ gives $n_e$=157\,542
degrees of freedom in (\ref{e2tt}). Fig.~\ref{Z21Trans} shows the associated
impedances for the frequency range $[f_{\rm min},f_{\rm
max}]=[1,7.5]$ GHz obtained using the conjugate gradient method when the ports are placed $2.5$ mm distant from
the PEC walls.

\begin{figure}[htbp]
    \begin{center}
    \subfigure[]{\includegraphics*[width=0.49\columnwidth]{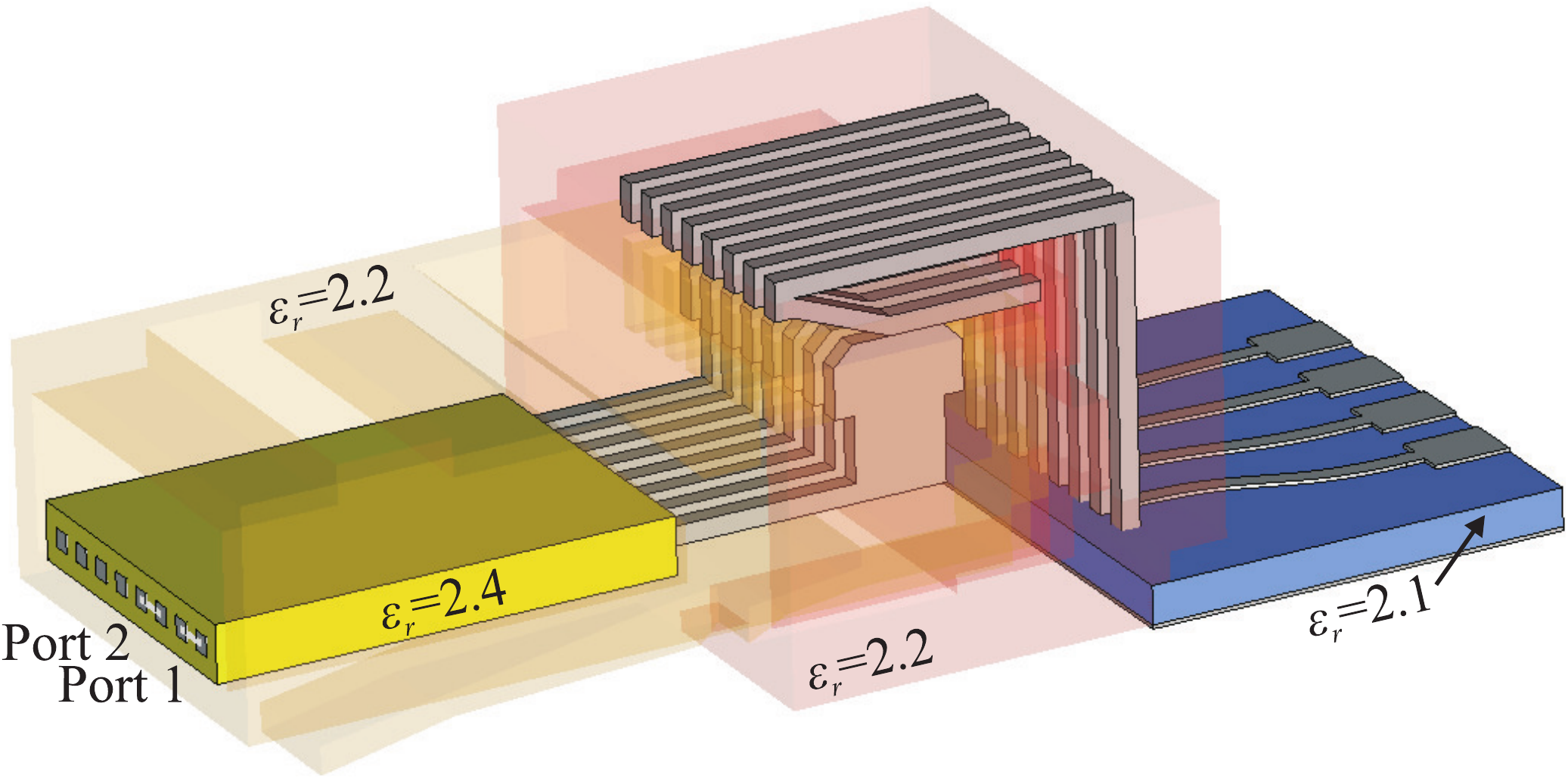}\label{Connector}}
    \subfigure[]{\includegraphics*[width=0.49\columnwidth]{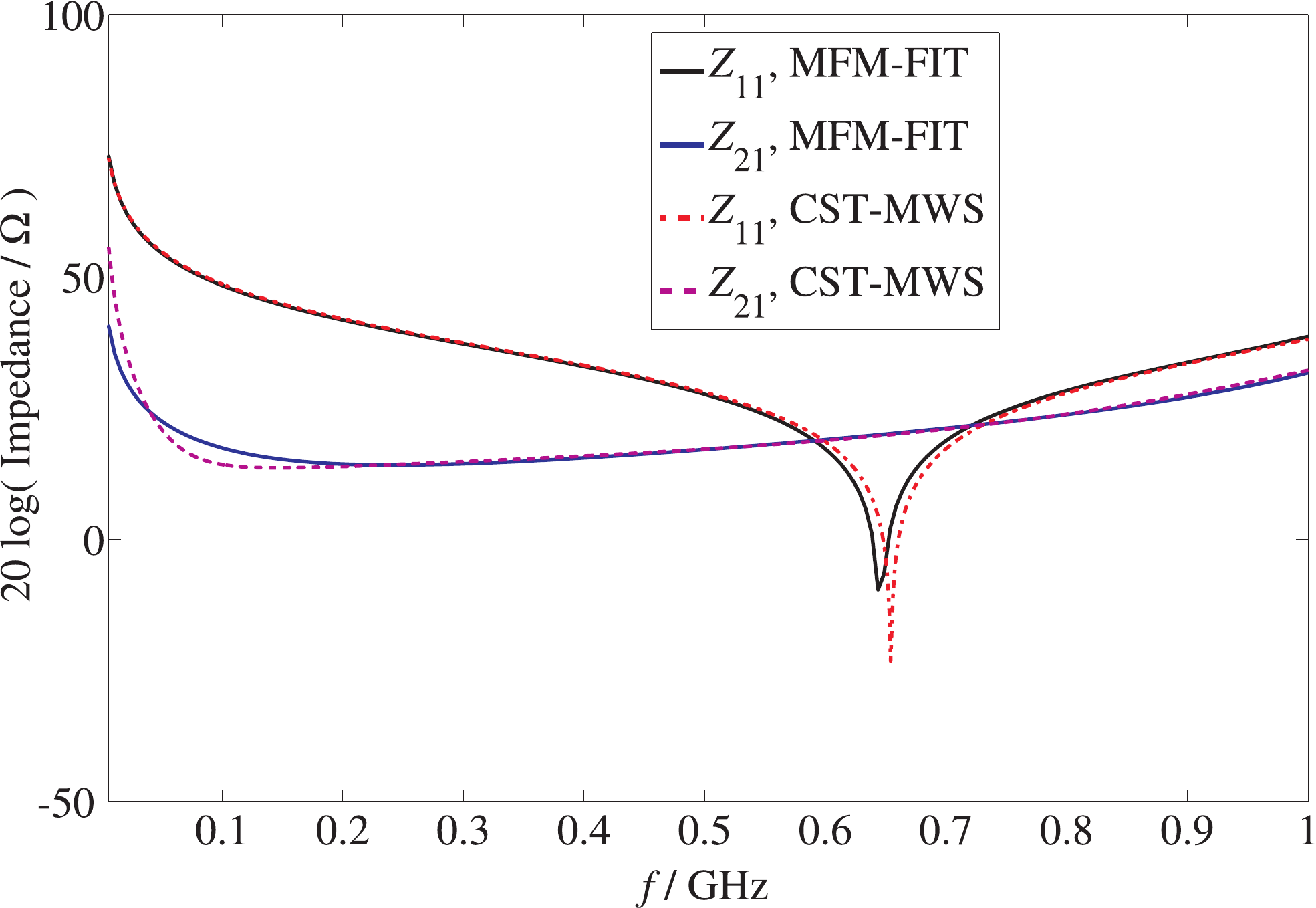}\label{ZConnector}}
    \end{center}
\caption{(a) An RJ45 connector containing 4 pairs of wire conductors and the plastic socket. The model dimensions can be found in CST STUDIO SUITE$^{\,\rm TM}$ examples. (b) The input impedance $Z_{11}$ and the crosstalk impedance $Z_{21}$ for the connector RJ45 in Fig.~\ref{Connector}.}\label{}
\end{figure}

The next case study is composed of several different materials. An
RJ45 connector, consisting of a socket jack and 4 differential pairs
of wires for the signal transmission, is analyzed. The wires of the
male socket are fixed to a substrate plate, the other wires are
connected to a metallic ground plane for shielding purposes,
Fig.~\ref{Connector}. The thickness of the metal connectors are
$0.5$ mm, the substrate thickness is $1$ mm, and the insulator grip
thickness is $1.5$ mm. The simulation domain is confined to a
$34\times14\times11.75~\rm mm^3$ metallic box. The generated mesh
$N_x$=63, $N_y$=45, $N_z$=28 yields to $n_e$=238\,140 degrees of
freedom in (\ref{e2tt}) which are solved using the KSP conjugate residual method. The discrete port 1 imposes a unit current as the input at
the plug end of the connector and the crosstalk voltage along the
coupled port 2 at the same side of the connector is shown in
Fig.~\ref{ZConnector}.

\begin{figure}[htbp]
    \begin{center}
    \subfigure[]{\includegraphics*[width=0.49\columnwidth]{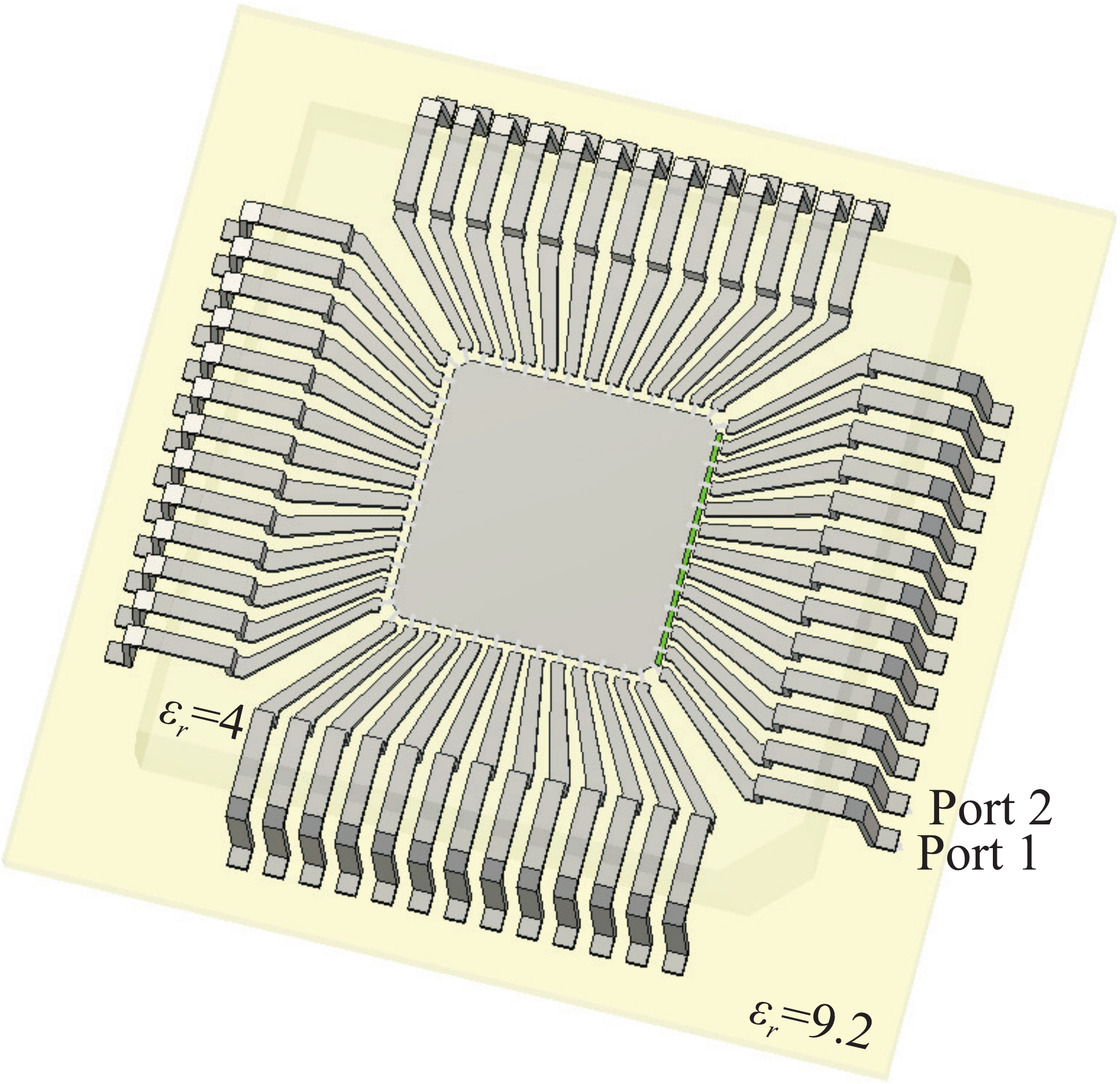}\label{ICpackage}}
    \subfigure[]{\includegraphics*[width=0.49\columnwidth]{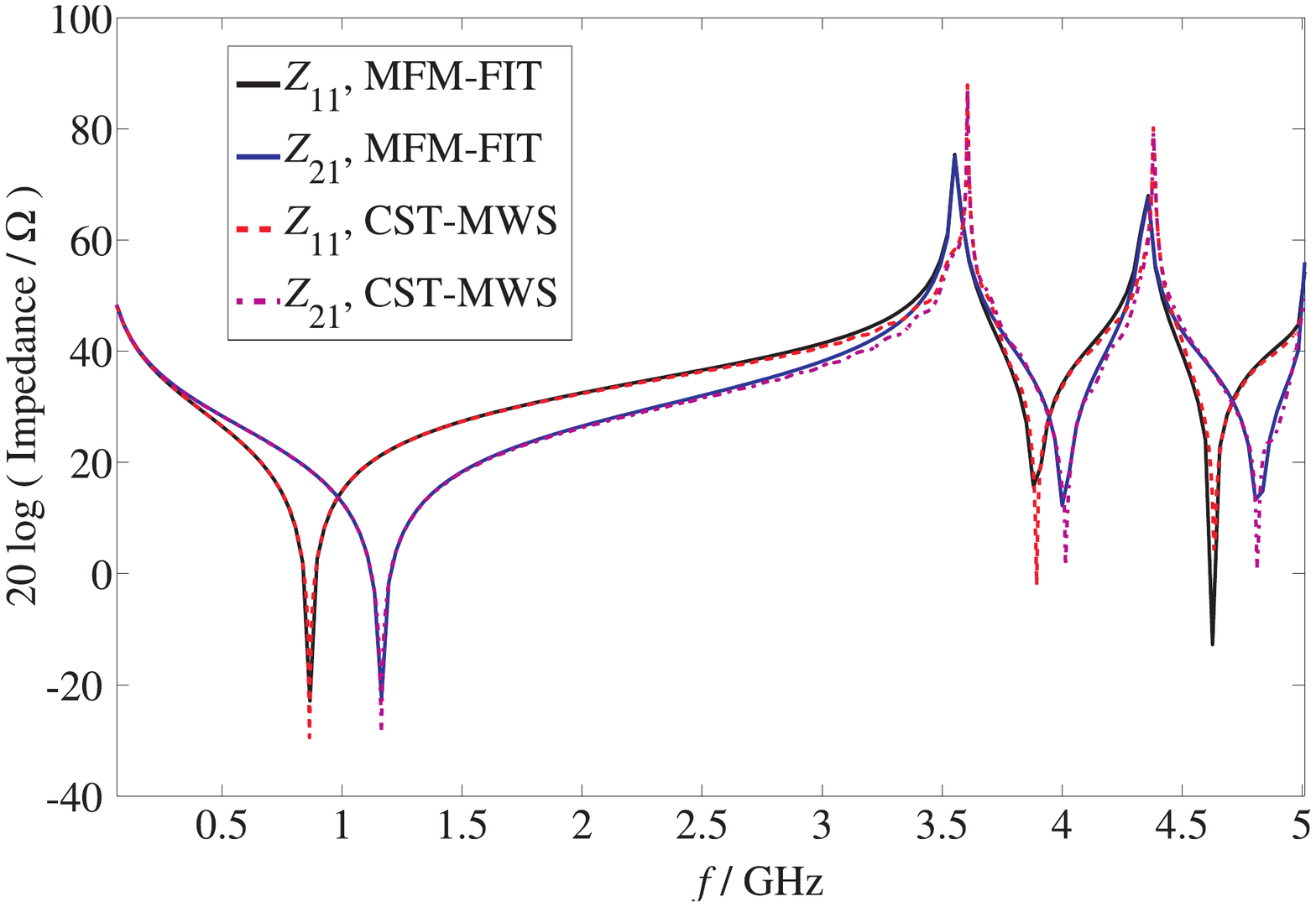}\label{Z21ICpackage}}
    \end{center}
\caption{(a) A 52-pin IC package consisting of a silicon chip encapsulated within a nonconductive plastic compound. The model size can be found in CST-MWS examples \cite{MWS}. (b) The input impedance $Z_{11}$ and the coupling impedance $Z_{21}$ for the IC package in Fig.~\ref{ICpackage}.}\label{}
\end{figure}

As the final example, a quad flat IC package consisting of a silicon
chip with the relative permittivity $\epsilon_r$=12.3 encapsulated
within a nonconductive plastic compound is simulated. The surface
mounted package is concentrically soldered to the PCB with a PEC
ground layer, Fig.~\ref{ICpackage}. The thickneParallelss of the metal
strips is $0.1$ mm, the substrate thickness is $0.5$ mm, and the
plastic thickness is $2.1$ mm. The electric BC is considered for all
sides of the $13.4\times13.4\times3.075968~\rm mm^3$ surrounding
box. The structured mesh $N_x$=65, $N_y$=65, $N_z$=14 results in
$n_e$=177\,450 degrees of freedom. The discrete port 1 imposes a
specific current as the input and (\ref{e2t}) is solved for
$N_f$=200 samples in the frequency range $[f_{\rm min},f_{\rm
max}]=[0.06,6]$ GHz using the biconjugate gradient stabilized method. The induced voltage along the port 1 and the
coupled voltage along the neighboring pad are read as the outputs.
The broadband results shown in Fig.~\ref{Z21ICpackage} comply with
the CST-MWS transient solver results on the same hexahedral mesh. The enormous abstraction of the code in PETSc environment eases the solver modification, for example when mutual field coupling between many ports in Fig.~\ref{ICpackage} are to be calculated. In fact, one only needs to update the excitation vector on the right hand side of (\ref{e2}) or (\ref{e2s}), likewise (\ref{e2t}) or (\ref{e2tt}), by activating all pins of the chip at single run and the solver has not to be run for each single port separately. 
\vspace{-1.5cm}





\section{Conclusions}

The matrix-free methods were adopted to the FIT scheme to minimize
the number of multiplications and memory requirements in the
construction of the final system of linear equations. The matrix-free FIT paves the path for realistic modeling of electrically large-scale radiation problems. The introduced recipe provides extreme ease of modifications in the kernel of the
applied algorithm, when material tensors are diagonal which is the case in most practical applications with isotropic materials. Initial index reordering of the unknowns was
applied to make the FIT system matrix diagonally dominant. The
rearrangement also facilitated the decomposition of large domain
into slices for passing the mesh information to different machines
in the pre-processing stage. Using the PETSc framework for
high-performance computing, the accuracy and efficiency of different
KSP solvers on the shell matrix were investigated. The biconjugate gradient stabilized and the conjugate
residual methods are shown to be optimal with respect to the solution
time among the other alternative iterative solvers for the presented problems with different sizes. Performing large-scale simulations with 1.5 billion unknowns
on 64 processors takes about several hours per frequency sample with
6 GB peak memory usage. The method also permitted running 102
millions of unknowns on a shared-memory multiprocessor system in
less than half a day.


\section*{Acknowledgement}
This work was initiated under the German Federal Ministry of
Education and Research (BMBF) funds to the MoreSim4Nano partner in
TU Darmstadt with contract 03MS613G.

\bibliographystyle{IEEEtran}
\bibliography{BibMoreSim4Nano}

\begin{thebibliography}{10}
\providecommand{\url}[1]{#1}
\csname url@samestyle\endcsname
\providecommand{\newblock}{\relax}
\providecommand{\bibinfo}[2]{#2}
\providecommand{\BIBentrySTDinterwordspacing}{\spaceskip=0pt\relax}
\providecommand{\BIBentryALTinterwordstretchfactor}{4}
\providecommand{\BIBentryALTinterwordspacing}{\spaceskip=\fontdimen2\font plus
\BIBentryALTinterwordstretchfactor\fontdimen3\font minus
  \fontdimen4\font\relax}
\providecommand{\BIBforeignlanguage}[2]{{%
\expandafter\ifx\csname l@#1\endcsname\relax
\typeout{** WARNING: IEEEtran.bst: No hyphenation pattern has been}%
\typeout{** loaded for the language `#1'. Using the pattern for}%
\typeout{** the default language instead.}%
\else
\language=\csname l@#1\endcsname
\fi
#2}}
\providecommand{\BIBdecl}{\relax}
\BIBdecl

\bibitem{Lee11}
Y.~Shao, Z.~Peng, and J.~F. Lee, ``Full-wave real-life 3-{D} package signal
  integrity analysis using nonconformal domain decomposition method,''
  \emph{{IEEE} Trans. Microw. Theory Tech.}, vol.~59, no.~2, pp. 230--241, Feb.
  2011.

\bibitem{Mittra11}
W.~Yu, X.~Yang, Y.~Liu, and R.~Mittra{\it~et al.}, ``New development of
  parallel conformal {FDTD} method in computational electromagnetics
  engineering,'' \emph{{IEEE} Antennas Propag. Mag.}, vol.~53, no.~3, pp.
  15--41, 2011.

\bibitem{Lavranos09}
C.~S. Lavranos and G.~A. Kyriacou, ``Eigenvalue analysis of curved waveguides
  employing an orthogonal curvilinear frequency-domain finite-difference
  method,'' \emph{{IEEE} Trans. Microw. Theory Tech.}, vol.~57, no.~3, pp.
  594--611, Mar. 2009.

\bibitem{Gdansk01}
P.~Przybyszewski, ``Fast finite difference numerical techniques for the time
  and frequency domain solution of electromagnetic problems,'' Supervised by M.
  Mrozowski, Technical University of Gdansk, 2001.

\bibitem{Schuhmann01}
R.~Schuhmann and T.~Weiland, ``Conservation of discrete energy and related laws
  in the finite integration technique,'' \emph{Progress In Electromagnetics
  Research}, vol.~32, pp. 301--316, 2001.

\bibitem{Erion07}
E.~Gjonaj, T.~Weiland, I.~Munteanu, and P.~Thoma, ``A parallel electromagentic
  simulation approach for the signal integrity analysis of {IC} packages,'' in
  \emph{Proc. {IEEE} Int. Electromagn. Compat. Symp. (EMC'07)}, Honolulu, {HI},
  2007, pp. 1--5.

\bibitem{Ren13}
T.~Ren, T.~Kalscheuer, S.~Greenhalgh, and H.~Maurer, ``Boundary element
  solutions for broadband 3{D} geo-electromagentic problems accelerated by
  multi-level fast mutlipole method,'' \emph{Int. J. Geophys.}, vol. 192,
  no.~2, pp. 473--499, Feb. 2013.

\bibitem{Parallel12}
T.~Iwashita, Y.~Hirotani, T.~Mifune, T.~Murayama, and H.~Ohtani, ``Large-scale
  time-harmonic electromagnetic field analysis using a multigrid solver on a
  distributed memory parallel computer,'' \emph{Parallel Computing}, vol.~38,
  no.~9, pp. 485--500, Sep. 2012.

\bibitem{PETSc}
S.~Balay, J.~Brown, K.~Buschelman, V.~Eijkhout, W.~Gropp, D.~Kaushik,
  M.~Knepley, L.~C. McInnes, B.~Smith, and H.~Zhang, \emph{{PETSc Users Manual}
  Revision 3.4}, Argonne, {IL}, May 2013.

\bibitem{MWS}
``{CST STUDIO SUITE$\;^{\rm TM}$} 2013,'' {CST - Computer Simulation Technology
  AG}, Bad Nauheimer Str. 19, 64289 Darmstadt, Germany, Tech. Rep., 2013.

\bibitem{Clemens01}
M.~Clemens and T.~Weiland, ``Discrete electromagnetism with the finite
  integration technique,'' \emph{Progress In Electromagnetics Research},
  vol.~32, pp. 65--87, 2001.

\bibitem{Rumpf14}
R.~C. Rumpf, C.~R. Garcia, E.~A. Berry, and J.~H. Barton, ``Finite-difference
  frequency-domain algorithm for modeling electromagnetic scattering from
  general anisotropic objects,'' \emph{Progress In Electromagnetics Research
  B}, vol.~61, pp. 55--67, 2014.

\bibitem{Shin13}
W.~Shin, ``3{D} finite-difference frequency-domain method for plasmonics and
  nanophotonics,'' Ph.D. dissertation, Stanford University, 2013.

\bibitem{Tijhuis08}
A.~Chabory, B.~de~Hon, W.~Schilders, and A.~Tijhuis, ``Preconditioned
  finitedifference frequency-domain for modelling periodic dielectric
  structures - comparisons with fdtd,'' in \emph{Proc. 38th European Microwave
  Conf. (EuMC'08)}, Amsterdam, the Netherlands, Oct 2008.

\bibitem{Hager11}
G.~Hager and G.~Wellein, \emph{Introduction to High Performance Computing for
  Scientists and Engineers}, 1st~ed.\hskip 1em plus 0.5em minus 0.4em\relax CRC
  Press, Taylor \& Francis Group, 2011.

\end{thebibliography}

\clearpage

\end{document}